\documentstyle[aps]{revtex}
\begin{document}
\draft
\title
      {
        Molecular dynamics simulation of the fragile glass former 
        ortho-terphenyl: a flexible molecule model
      }
\author{        
        S.~Mossa$^{1}$,
        R.~Di Leonardo$^{1}$,
        G.~Ruocco$^{1}$,
        M.~Sampoli$^{2}$
                }
\address{
         $^1$
         Dipartimento di Fisica and INFM, Universit\`a di L'Aquila,
         Via Vetoio, Coppito, L'Aquila, I-67100, Italy \\
         $^2$
         Dipartimento di Energetica and INFM, Universit\`a di Firenze,
         Via Santa Marta 3 , Firenze, I-50139, Italy \\
        }

\date{\today}

\maketitle
\begin{abstract}

We present a realistic model of the fragile glass former 
orthoterphenyl and the results of extensive molecular dynamics
simulations in which we investigated its basic static and dynamic 
properties. In this model the internal molecular interactions 
between the three rigid phenyl rings
are described by a set of force constants, including harmonic
and anharmonic terms;
the interactions among different molecules are described by
Lennard-Jones site-site potentials.
Self-diffusion properties are discussed in detail together with
the temperature and momentum dependencies  
of the self-intermediate scattering function.
The simulation data are compared with existing experimental results
and with the main predictions of the Mode Coupling Theory.
\end{abstract}

\pacs{PACS numbers: 64.70.Pf, 71.15.D, 61.25.E, 61.20}

\section{INTRODUCTION}
\label{introduction}

In the recent years a renewed interest on the
glass transition phenomenon has motivated
extensive experimental and theoretical works
(see~\cite{angell} and reference therein). On
the theoretical side, new descriptions of the glass transition
has been developed: they emphasized
either the dynamic (as the Mode Coupling Theory (MCT)
of G\"{o}tze~\cite{mct1,mct2} or the coupled oscillators model of Ngai
and Tsang~\cite{ngaitsang})
or the thermodynamic (as the first principle computation
based on a replica formulation of~\cite{mezpar} or the 
inherent structure formalism computation of~\cite{sciorkobtar})
aspects of the transition itself.

A common feature of all these theories is they have
been developed for ``model systems'', often monoatomic models.
The comparison of the theoretical results with the 
real experiment are, therefore, complicated by the trivial observation that
in the real world the glass forming systems are made out
of ``molecular systems''. As a consequence in the current
literature there is a large debate on the applicability of the
theoretical predictions to the experimental outcome
and the more stringent tests of the theories come from
Molecular Dynamics (MD) works.

As an example, it is highly debated in literature the origin, 
in molecular glassformer, of
{\it secondary} relaxations 
that can be observed by several experimental techniques beside
the well-known microscopic and structural dynamics 
(see, among others, \cite{monfiomasc} and reference therein).
For instance, {\it fast} relaxations (i. e. in the $10^{-12}$ s range)  
have been observed in several glasses and it is not yet clear
if their origin is related to the molecular center of mass motion, 
as MCT would explain in terms of
$\beta$-process, 
or rather to rotational or intramolecular dynamics.   
It is clear the crucial role that 
MD simulations can play to solve this specific problem. 
If it is possible to build 
a {\it realistic} model able to take 
into account the internal degrees of freedom,
as well as the translational dynamics, computer simulations allow one to
access any observable quantity of the system, also
those not directly measurable by present
experimental techniques. Such possibility, together with
the physical intuition, could allow the
identification of the microscopic mechanisms underlying the different
observed relaxation processes.

In this paper we want to address the problem to set up
a ``realistic'' potential for a glass model system capable 
to account for the internal
molecular degrees of freedom.
Among the glassforming molecular liquids
characterized by an extremely rich dynamical behavior,
the organic {\it fragile}~\cite{angell} glass former {\it ortho-terphenyl} (OTP) 
($T_{m}=329$ K, $T_{c}\simeq 290$ K~\cite{petbarpuj}, $T_{G}=243$ K)
has received much attention from both experimental and
numerical simulation points of view. 
The structure of {\mbox OTP} molecule, shown in Fig.~\ref{OTP},
is known from neutron~\cite{bus34} and X-ray~\cite{bus35} 
diffraction studies; in condensed state the OTP
molecules are bound together only 
by Van der Waals forces, which resemble 
the Lennard-Jones ones often used by most
theories and computer simulations aiming to study the 
glass transition problem.

Due to its structural complexity, if we would like to 
obtain reliable results to be compared
with experimental data, we need to take into
account not only the translations of the molecular center of masses
and the rotations of the molecules as a whole, but also
intramolecular motions like stretching along the molecular bonds,
tilt of the bonds, rotation of the side rings with respect to the central one
and so on. In other words, we need to describe the dynamics of the
liquid at {\it atomic} level. On the contrary, in order to set up
a computational scheme which is affordable in
not-too-long time with the nowadays computer capability,
we need the {\it simplest} model potential able to capture
the relevant features of the 
dynamical behavior of the real system.

In the literature numerical studies of OTP have been proposed
making use of different techniques ranging from {\it harmonic lattice
dynamics}~\cite{criaber} to molecular dynamics simulations on atomic level
based on a general force field provided by the standard program
Alchemy III~\cite{ouchen}. Nevertheless, at our knowledge, 
only two studies based on molecular dynamics simulations
of {\it molecular models} of OTP have been proposed so far:

{\it i) Lewis et al.} \cite{lewa,rolngai} 
represent the  molecule like a 
three-sites complex, each site playing the role of a whole phenyl ring,
without internal dynamics and an intermolecular interaction of the 
Lennard-Jones (LJ) type. This model take in account only 
the dynamical behavior associated with the translations of the molecular 
center of masses and with the rotations of the molecules as a whole;

{\it ii) Kudchadkar and Wiest} \cite{kuwi} propose a more realistic model
with the ``true'' structure of the molecule. The intermolecular interaction
is of LJ type and, as internal degrees of freedom, 
only the rotational dynamics of the side rings with respect to the 
central one is taken into account. 
These internal degrees of freedom are effectively the most relevant,
nevertheless the authors parameterize the potential 
in such a way that the side rings are, at equilibrium,
in a configuration that corresponds to a saddle
point in the molecular energy surface.

The model potential we are going to introduce 
results to be much more efficient 
in mimicking the complexity of the dynamical behavior of 
the real system.
The paper is organized as follows: in the next section we introduce
the intramolecular model potential; in 
section~\ref{force constants} we explain how
we calculated the force constants in order to reproduce a realistic
isolated molecule vibrational spectrum. 
In section~\ref{computation} we present some computational
details and in section~\ref{MCT}
we remind some of the main well-established predictions of the 
ideal Mode Coupling Theory used in section~\ref{results} to test
the center of mass dynamical behavior; 
In section~\ref{results} we discuss our MD simulation results
mainly with regard to the study of the diffusion and self-dynamic properties. 
The last section contains an overall discussion and the conclusions.

\section{THE MOLECULAR MODEL}
\label{molecular model}
In our model the OTP molecule is constituted by three {\it rigid} hexagons 
(phenyl rings) of side $L_{a}=0.139$ nm  connected as shown in Fig.~\ref{OTP},
i.~e. two adjacent vertices of the {\it parent} (central) 
ring are bonded to one vertex of
the two {\it side} rings by bonds whose length, at equilibrium, is 
$L_{b}=0.15$ nm .
In our scheme, each vertex of the hexagons is thought to be occupied
by a fictious atom of mass $M_{CH}=13$ a.m.u. representing a carbon-hydrogen 
pair (C-H). The choice of such a fictious atom, with its renormalized mass, greatly
simplify the computer simulation but presents some drawbacks: {\it i)} in the 
real molecule the two couples of carbon atoms connecting the rings are not 
bonded to hydrogen atoms, while in our model we consider all the 18 vertices 
having the same mass $M_{CH}$ so that the total molecular mass is 
overestimated (234 rather than 230 a.m.u) and {\it ii)} 
the moments of inertia of the model 
rings are smaller than the real ones the hydrogen atoms being too 
close to the ring center. Nevertheless, we expect only minor effects 
on the overall dynamics from the previous simplification.

The three rings of a given molecule interact
among themselves by an {\it intramolecular} potential, such potential being
chosen {\it i}) to preserve the molecule from "dissociation";
{\it ii}) to give the correct relative equilibrium positions for the three
rings; {\it iii}) to represent the real
intramolecular vibrational spectrum as close as possible.
The interaction among different molecules, actually among the rings 
pertaining to different molecules, is accounted for by a site-site 
pairwise additive potential energy of the (6-12) Lennard-Jones type, 
each site being one of the hexagons vertices.

To sum up, the total interaction potential energy is
written as the sum of an {\it intermolecular} and an {\it intramolecular} term:
\begin{equation}
V_{tot}=V_{inter}+V_{intra}.
\end{equation}
The first term can be written explicitly as:
\begin{equation}
V_{inter}= \frac{1}{2} \sum_{i \neq j} \sum_{\xi \xi'} \sum_{\ell \ell'} 
           V_{LJ} (\vert \bar r_{i\xi\ell} - \bar r_{j\xi'\ell'} \vert)
\end{equation}
where $\bar r_{i\xi\ell}$ is the position of the $\ell$-th atom ($\ell$=1\ldots6)
in the $\xi$-th ring ($\xi$=1 \ldots 3, hereafter $\xi$=1 indicates 
the parent ring) belonging to
the $i$-th molecule ($i$=1\ldots N), and
\begin{equation}
  V_{LJ}(R)=4\epsilon\left[ \left(\frac{\sigma}{R}\right)^{12}-
             \left(\frac{\sigma}{R}\right)^{6}\right].
\nonumber
\end{equation}
The optimal choice of the two intermolecular force parameters,
$\epsilon$ and $\sigma$, will be discussed later.

In principle, the {\it intramolecular} interaction potential can be expressed 
in terms of the degrees of freedom describing the center of mass positions 
($\bar R_{\xi}$, $\xi$=2,3) and orientations (e.g. the set of Eulerian angles) 
of the side rings with respect to the parent ring.
However, for computational purposes,
it is simpler to express the intramolecular potential in terms of 
orthonormal unit vectors attached to each ring or better in terms of quantities
built from these vectors. 
With reference to Fig.~\ref{geometry} the sets of unit vectors for each ring 
$\left\{\hat{l}_{\xi},\hat{m}_{\xi},\hat{n}_{\xi}\right\}$ are defined as

\begin{eqnarray*}
\hat{l}_{\xi}          & = & (1,0,0)             \\ 
\hat{m}_{\xi}          & = & (0,1,0)             \\
\hat{n}_{\xi}          & = & (0,0,1)             
\end{eqnarray*}

i.e. $\hat{l}_{\xi}$ and $\hat{m}_{\xi}$ are orthogonal unit vectors in the 
ring plane, while $\hat{n}_{\xi}$ is the normal to that plane.

The unit vectors
that are parallel to the ring-ring bonds at equilibrium, are given by
\begin{eqnarray*}
\bar{u}_{2} &=&\frac{1}{2}\left(\hat{l}_{2}+\sqrt{3}\hat{m}_{2}\right) \\
\bar{u}_{3} &=&\frac{1}{2}\left(-\hat{l}_{3}+\sqrt{3}\hat{m}_{3}\right)\\
\bar{u}_{1(2)}&=&\frac{1}{2}\left(\hat{l}_{1}+\sqrt{3}\hat{m}_{1}\right) \\
\bar{u}_{1(3)}&=&\frac{1}{2}\left(-\hat{l}_{1}+\sqrt{3}\hat{m}_{1}\right);
\end{eqnarray*}

The positions of the four Carbon atoms that
link the three rings, i.e $\bar P_{1(2)}$ and $\bar P_{1(3)}$ 
in the parent ring
and $\bar P_2$ and $\bar P_3$ in the side rings, are given by,
with respect to their ring centers $\bar R_{\xi}$,  
$\bar P_{1(2)}-\bar R_1=L_a \bar u_{1(2)}$, $\bar P_{1(3)}-\bar R_1=L_a \bar u_{1(3)}$,
$\bar P_{2}-\bar R_2=L_a \bar u_{2}$ and $\bar P_{3}-\bar R_3=L_a \bar u_{3}$.

Finally, it is useful to define four further interaction sites
two pertaining to the parent ring,
$\bar S_{1(2)} = \bar R_1 + L_{c} \bar u_{1(2)} $  and 
$\bar S_{1(3)} = \bar R_1 + L_{c} \bar u_{1(3)} $
and two pertaining to the side rings (2 and 3 respectively),
$\bar S_{2} = \bar R_2 - L_{c} \bar u_{2} $ and 
$\bar S_{3} =\bar R_3 - L_{c} \bar u_{3} $.
Here $L_{c}=L_a+L_b/2$, so that at the equilibrium position 
$\bar S_{1(2)} \equiv \bar S_{2} $ and  $\bar S_{1(3)} \equiv \bar S_{3} $.

The variable we have introduced will be used to simply express 
the different contributions to the intramolecular potential 
in the next subsections.

\subsection{Stretching along the ring-ring bonds and between the side rings}
The fluctuations of the distances among the centers of mass 
of the three rings are accounted for by introducing three
``springs''. The parent-side ring stretching implies the elongation of
a C-C bond, which is expected to have a high stiffness.
The corresponding potential in harmonic approximation is written as:
\begin{eqnarray}
V_{S}&=& c_{1} \left [\vert {\bar P}_{1(2)}-{\bar P}_{2}\vert -L_b \right ]^{2}+\\
&&c_{1} \left [ \vert {\bar P}_{1(3)}-{\bar P}_{3}\vert -L_b \right ]^{2}
\nonumber
\end{eqnarray}
On the other hand, no direct chemical bond is present between the two side rings, 
and we expect a less stiff spring for the fluctuation of the distance between the
side ring centers. We model this interaction by:
\begin{equation}
V_{B} = c_{2} \left [ \vert {\bar R}_{2}-{\bar R}_{3}\vert -(2L_c) \right ]^{2}
\end{equation}
The determination of the force constants $c_1$ and $c_2$, 
as well as the others we are going to introduce, will be discussed later.

\subsection{Tilt of the ring-ring bond}
In the OTP crystal structure\cite{busing}, the bond angles
$\bar P_2-\bar P_{1(2)}-\bar P_{1(3)}$ and \mbox{$\bar P_3-\bar P_{1(3)}-\bar P_{1(2)}$}
are 123.6$^o$ and 123.0$^o$ respectively, while the angles
$\bar P_2-\bar P_{1(2)}-\bar P_{A}$ and $\bar P_3-\bar P_{1(3)}-\bar P_{B}$
are 118.4$^o$ and 117.4$^o$ (see Fig.~\ref{geometry})
Further, in the isolated molecule, the ring-ring bonds
are forced out of the plane of the parent ring so that the dihedral angle
\mbox{$\Phi=\bar P_2-\bar P_{1(2)}-\bar P_{1(3)}-\bar P_{3}$} is 5.2$^o$. 
This lack of planarity is due to the little asymmetry introduced 
by the difference between a Carbon bonded to an Hydrogen and 
a Carbon bonded to a Carbon of another ring.
In our model, all these angles at the equilibrium are put equal to 120$^o$
and $\Phi$ equal to $0$. 

We model the restoring forces for these angles
by using the scalar product of the unit vectors $\bar u_2$ and $\bar u_{1(2)}$
(as well as that of $\bar u_3$ and $\bar u_{1(3)}$). Being
\mbox{${\bar u}_{2}\cdot {\bar u}_{1(2)}={\bar u}_{3}
\cdot {\bar u}_{1(3)}= 1$}, 
at equilibrium, the quadratic term in 
the small oscillation approximation is given by:
\begin{equation}
V_{T_1} = c_{3} (1 - {\bar u}_{2}\cdot {\bar u}_{1(2)}) + 
          c_{3} (1 - {\bar u}_{3}\cdot {\bar u}_{1(3)})  
\end{equation}
However, this term is not enough to ensure the co-planarity of the 
vectors $\bar u_2$ and $\bar u_{3}$ with the parent ring; to force the rings 
towards the
co-planar equilibrium condition we make use of the ``sites'' $\bar S_2$ and 
$\bar S_{1(2)}$ (as well as $\bar S_3$ and  $\bar S_{13}$) introducing 
between them a spring of vanishing equilibrium length:
\begin{eqnarray}
V_{T_2}&=& c_{4} \vert {\bar S}_{1(2)}-{\bar S}_{2}\vert^{2}+\\
&&c_{4} \vert {\bar S}_{1(3)}-{\bar S}_{3}\vert^{2}\nonumber
\end{eqnarray}

\subsection{Rotation of the side rings along the ring-ring bond}
Inside the intramolecular dynamics, 
we expect that a crucial role is played by the rotation of the side rings planes 
around the ring-ring bonds\cite{busing,werall} in interfering or modifying 
the intermolecular relaxation processes. 
The relevant variables 
describing this motion are the two angles $\left\{ \phi_{2},\phi_{3} \right\}$ 
between the normals to the side ring and parent ring planes.
In the crystalline structure, the two side rings have slightly different angles, 42.4$^o$ and 62.0$^o$ \cite{busing}. 
However we model the disordered condensed phases (liquid and glass) 
by using as equilibrium angle values those of the isolated molecule, i.e. 54.0$^o$,
as discussed below.
We have to remark that the isolated molecule symmetry implies 
two iso-energetic configurations separated by a finite barrier
as is qualitatively illustrated in Fig.~\ref{map}.
We performed an {\it ab-initio} calculation of the single molecule
potential energy surface
as a function of $\phi_{2}$ and $\phi_{3}$ with all the other internal degrees of 
freedom fixed to their equilibrium values. Such calculation consists in 
the minimization of the Hartree-Fock energy over the gaussian
basis set  3-21G. For each atomic species the inner shell is made up of
3 gaussians while the valence shell is a linear combination
of 2 gaussians orbital plus a 1 gaussian orbital; 
the minimization was carried out by the standard package Gaussian 94.
In Fig.~\ref{map} a contour plot of the Hartree-Fock 
potential energy  is shown as a function of the rotation angles.
It is important to quote that this map has only a 
semi-quantitative meaning, the structure of the whole molecule
being not re-optimized during the scan of $\phi_{2}$ and $\phi_{3}$ angles. 
However, a careful study, performed reoptimizing the whole molecular
structure, has been done around the saddle point
$\phi_{2}=\phi_{3}=90^{o}$ and around the two equivalent minima 
that turn out to be at $\phi_{2}=\phi_{3}=54^{o}$ 
and $\phi_{2}=\phi_{3}=126^{o}$.
From this calculation the barrier height has been estimated 
to be $V_s/k_B=580\,K$ 
and, from this value, it is possible to envisage the nature of
the rotational motion at the temperatures of interest: the two side rings can 
pivot in phase around the bonds crossing from one minimum to the other 
degenerate one. Moreover they can perform
librational out-of-phase motions of (approximatively) harmonic type.

In order to represent this potential surface we 
express the in-phase rotation of the
two side rings with a high-order (6th) polynomium and the out-of-phase
rotation in the harmonic approximation.
For this purpose we use as primary variables 
the scalar products $\bar n_1 \cdot \bar n_2$ and $\bar n_1 \cdot \bar n_3$
(their equilibrium value being 
$(\bar n_1 \cdot \bar n_2)_{eq}=(\bar n_1 \cdot \bar n_2)_{eq}
=\alpha_{0}=0.59$) and to
disentangle the in-phase from the out-of-phase motion, we introduce the
two variables:
\begin{eqnarray}
\alpha&=&\frac{{\bar n}_{1}\cdot{\bar n}_{2}+{\bar n}_{1}\cdot{\bar n}_{3}}{2}
\\
\beta&=&\frac{{\bar n}_{1}\cdot{\bar n}_{2}-{\bar n}_{1}\cdot{\bar n}_{3}}{2}
\nonumber
\end{eqnarray}
The final form of the "internal rotation" potential will be
\begin{equation}
V_{R}  =  V_{R_1}({\alpha})+V_{R_2}({\beta})
\label{torsion}
\end{equation}
with
\begin{eqnarray}
V_{R_1}({\alpha}) & = & b_{1}\alpha^{2}+b_{2}\alpha^{4}+b_{3}\alpha^{6}
\label{vr1}\\
V_{R_2}({\beta})  & = & c_{6}\beta^{2}                  
\label{rotation}
\end{eqnarray}

The parameters $b_1$, $b_2$ and $b_3$ describing the in-phase rotation
potential are derived according to the following procedure.
In the harmonic approximation, in proximity of the equilibrium 
position ${\alpha_o}$ for the scalar products, it must hold 
\begin{equation}
V_{R_1}({\alpha})  \approx  c_{5} ( \alpha - {\alpha_o} )^{2}; 
\end{equation}
and taking into account that the barrier height 
must be equal to $V_s$, we have the following conditions:
\begin{eqnarray*}
V_{R_1}  ({\alpha_o}) & = & - {V_s}  \\
V_{R_1}' ({\alpha_o}) & = & 0       \\
V_{R_1}''({\alpha_o}) & = & 2 c_{5} 
\end{eqnarray*}
implying
\begin{eqnarray}
b_{1} & = & -\frac{{V_s}}{{\alpha_o}^{2}} + \frac{c_{5}}{4} 
            +2 b_{3} {\alpha_o}^{4}                                                     \nonumber\\
b_{2} & = & \frac{c_{5}}{4 {\alpha_o}^{2}} - 3 b_{3} {\alpha_o}^2              \label{bval}     \\
b_{3} & = & \frac{1}{{\alpha_o}^{4}} \left( -\frac{{V_s}}{{\alpha_o}^{2}}
+\frac{c_{5}}{4} \right)                                                       \nonumber
\end{eqnarray}

\section{Force constants}
\label{force constants}
We can finally write our internal model potential like
\begin{equation}
V_{intra}=V_{S}+V_{B}+V_{T_1}+V_{T_2}+V_{R_1}+V_{R_2};
\end{equation}
this potential is parametrized by the set of six free coefficients $\{c_{k}\}$
whose actual value can be tuned in  order to obtain a {\it realistic free
molecule vibrational spectrum}. 
Indeed, in the small oscillations approximation,
we can determine the values of the coefficients $\{c_{k}\}$
by diagonalizing the 
dynamical matrix and fitting the resulting eigenfrequencies
$\omega^{\lambda}_{DIAG}$
to the {\it lowest} frequencies $\omega^{\lambda}_{HF}$ obtained by
an Hartree-Fock calculation 
of the vibrational frequencies in the electronic 
ground state of the isolated molecule.

We have $18$ eigenvalues for the dynamics of three
rings ($\lambda=1,\ldots 18$)
but actually 
only $N_{e}=12 $ eigenvalues $\omega^{\lambda}_{DIAG}$ are non vanishing, 
the other being associated with the translations  
rotations of the molecule as a whole.

More explicitly, the set $\{c_{k}\}$ is obtained by minimizing,
by the standard Levenberg-Marquardt algorithm\cite{minuit}, the error function
\begin{equation}
\sum_{\lambda=1}^{N_{e}}
(\omega^{\lambda}_{DIAG}-\omega^{\lambda}_{HF})^{2}
\label{chiquadro}
\end{equation}
where, for a given set of  $\{c_{k}\}$, the quantities
$\omega^{\lambda}_{DIAG}$ are the solution of the 
eigenvalues problem~\cite{goldstein}
\begin{equation}
|\hat{V} - \omega^{2}\hat{T}|=0.
\end{equation}
Here $\hat{V}$ and $\hat{T}$ are the hessian matrixes of the second partial
derivatives of the potential and kinetic energy respectively, with respect
translational and rotational degrees of freedom of the three rings.

In Tab.~\ref{coeff} are reported the values of the coefficients $\{c_{k}\}$ 
determined
by the minimization together with the values for $\{b_{k}\}$ 
derived from Eqs.~(\ref{bval}). 
The corresponding set of frequencies is shown in Tab.~\ref{frequencies}
where $\omega_{MD}$ are the frequencies derived from a computer simulation on a
system of isolated molecules at low temperature. 
These frequencies have been identified via the peaks 
of the spectrum of the velocity autocorrelation function.

Other model details are reported on ref.~\cite{phdste}.

\section{COMPUTATIONAL DETAILS}
\label{computation}

We have studied a system composed by 108 molecules 
(324 rings, 1944 interaction centers); the sample is
large enough to neglect finite size effects on the
investigated properties with reasonable computation times.
The values of the parameters entering the site-site 
Lennard-Jones interactions 
were determined by preliminary simulations: the value of
$\epsilon$ were firstly determined by comparing the computed
and the experimental~\cite{calldougfal} self-diffusion 
coefficients versus temperature
in the range $380$ K $\leq T \leq 440$ K; the value of $\sigma$
was estimated by tuning the static structure factor so to place 
the first maximum in the right position~\cite{barbertchi}. 
Successive iterations led to 
$\epsilon / k_{B} =14$ K and $\sigma=0.4$ nm.
To speed up the calculation of the intermolecular potential
and to assure all the torques to be estimated in a consistent way,
the cut off for L-J interactions is applied between
rings centers, i.e. the sites are considered not interacting
when they pertain to rings whose center distance is larger than
$r_{c}\simeq 3 (\sigma +  L_{a}) = 1.6 \, nm$. 

At this stage a remark about the rotational potential expressed in 
Eq.~(\ref{torsion})
is needed: such potential presents a serious drawback since 
the intrinsic ambiguity due to the parity of the scalar products 
involved can force the side rings in the wrong positions.
This problem can be bypassed introducing a new term 
in the intramolecular potential that 
gives zero contribution to the energy when the molecule is 
close to the equilibrium position. If we define the product 
$w = ({\hat l}_{1}\cdot {\hat n}_{2})({\hat l}_{1}\cdot {\hat n}_{3})$,
and we put $c_{7}=10\,c_{6}$ ( the actual value of $c_{7}$ is irrelevant,
it must only be able to force the rings in the right way ) we can write
\begin{equation}
V_{A}=\left\{
\begin{array}{ll}
c_7 w^2 & \mbox{if $w>0$}\\
0       & \mbox{otherwise}
\end{array}
\right.
\end{equation}
This term has been activated only during the preparation 
of the initial configuration (i.~e. at high temperature);
during the thermalized evolution of the system 
we expect molecules do not drive too much away 
from their equilibrium configuration
and then this term to be always equal to zero.
   
To integrate the equations of motion we have treated each ring 
as a separate rigid body, identified by the position  
of its center of mass $\bar R_{i\xi}$ and by its 
orientation expressed in terms 
of quaternions $\bar q_{i\xi}$ \cite{allen}.
The standard Verlet leap-frog algorithm \cite{allen} has been used
to integrate the translational motion while, for the most difficult
orientational part, the refined algorithm 
due to Ruocco and Sampoli~\cite{ruosam} 
has been employed; such choices allow a very stable integration
with a relatively long time step.

The rotational dynamical problem can be written as
\begin{eqnarray}
\label{rotmot}
\frac{d{\bar J}_{i\xi}}{dt}&=&{\bar\tau}_{i\xi}(\{{\bar R}_{j\xi'}\},
\{{\bar q}_{j\xi'}\})
\nonumber\\
&& \\
\frac{d{\bar q}_{i\xi}}{dt}&=&{\hat{\mathcal M}}
(\{{\bar q}_{i\xi}\})\cdot{\bar J}_{i\xi}\nonumber
\end{eqnarray}
where ${\bar J}_{i\xi}$ is the angular momentum of ring $\xi$
in the molecule $i$, ${\bar \tau}_{i\xi}$ the torque
acting on it and 
${\hat{\mathcal M}}_{i\xi}$ is the inverse of the inertia tensor 
in quaternion coordinates.

The expression of the torques is simplified as the
rotational part $V_{rot}$ 
of the intramolecular potential
energy has been expressed in terms of suitable scalar products written in the 
general form $s={\bar v}_{1}/\vert{\bar v}_{1}\vert
\cdot {\bar v}_{2}/\vert{\bar v}_{2}\vert$. 
Therefore the value of the torques $\tau_{v_1}$ and $\tau_{v_2}$
associated to the degree of freedom corresponding 
to the angle $\arccos(s)$ can then
be evaluated by \cite{goldstein}
\begin{equation}
\tau_{v_1}=-\tau_{v_2}=-\frac{\partial V_{rot}}{\partial s} ({\bar v}_{1}
\times {\bar v}_{2})
\end{equation}

The leap-frog algorithm for the rotational motion 
reported can be written in the form~\cite{ruosam}
\begin{eqnarray}
\label{rotation}
{\bar J}_{i\xi}(t+t')&=&{\bar J}_{i\xi}
\left(t-\frac{\Delta t}{2}\right)+
{\bar \tau}_{i\xi}(t)\left(\frac{\Delta t}{2}+t'\right)
\\
\mbox{with}&& 0\leq t' \leq \Delta t \nonumber 
\\
{\bar q}_{i\xi}(t+\tau)&=&{\bar q}_{i\xi}(t)+\int_{0}^{\tau} 
dt'\hat{{\mathcal M}}({\bar q}_{i\xi}(t+t'))
\cdot {\bar J}_{i\xi}(t+t')
\label{rot_alg}
\\
\mbox{with}&&0 \leq \tau \leq \Delta t \nonumber 
\\
{\bar J}_{i\xi}\left(t+\frac{\Delta t}{2}\right)&=&{\bar J}_{i\xi}
\left(t-\frac{\Delta t}{2}\right)+
{\bar \tau}_{i\xi}(t)\Delta t.
\end{eqnarray}
The crucial point is that the dependence of the matrix $\hat{{\mathcal M}}_{i\xi}$ on
the angular variables implies the need of using a time step $\Delta t'$ to perform 
the numerical integration appearing in Eq.~(\ref{rot_alg}),
smaller then $\Delta t$ used for the CMs integration.
In turn, the value of $\Delta t$ is limited by the highest vibrational frequency of about
$450\, cm^{-1}$. 
In our case, the chosen values of $\Delta t = 2$ fs and $\Delta t'=\Delta t/5$ are found to be sufficiently small to reduce the fluctuations
of the total energy to a negligible fraction of the kinetic energy. 
It is important to note 
that the use of the smaller $\Delta t'$ does not increase significantly the CPU computational time 
since the time consuming part in each MD step is the calculation of the 
forces and torques which are kept fixed during the integration of 
Eq.~(\ref{rotation}).

We considered a wide temperature range spanning the liquid phase and reaching
the region close to $T_{c}$ as shown in Tab.~\ref{protocol}
in which the whole set of simulation times is reported.
During the different temperature runs, the size 
of the cubic box has been rescaled in order to 
keep the system at the experimental density which,
for $T\geq T_{g}$, can be fitted by the polynomium \cite{giuphd}
\begin{equation}
\rho(T)= 
1.2983 - 7.00 \cdot 10^{-4} \cdot T - 1.23 \cdot 10^{-7} \cdot T^{2}
\end{equation}
where $\rho$ is in $g/cm^3$ and $T$ in K.
At each temperature $T$
we have organized the simulations following this scheme:
\begin{itemize} 
\item The system was coupled for a time
$t_{resc}$, chosen in a somehow arbitrary way, to a stochastic heat bath, 
i.e., the velocities of the rings were replaced following
a logarithmic pattern with the velocities 
drawn from a Boltzmann distribution corresponding to such temperature;
\item At this point the system was at the desired temperature and we let it to
perform a {\it microcanonical} time evolution (constant energy)
for a period $t_{term}$ comparable with the experimental 
structural relaxation time $\tau_{\alpha}$ at the same temperature;
in such a way we expect
every slow degree of freedom of the system to be correctly thermalized and we
control that there was no drift in temperature and 
the degree of energy conservation (fluctuations are always less than $1\%$
of the kinetic energy).
\item We considered the final system configuration obtained in this way
as a good equilibrium starting state for a molecular dynamics trajectory.
At each temperature we perform three different runs: the first one $20$ ps long
with a $4\times 10^{-3}$ ps configuration saving time has been used to compute
the small time dynamical behavior of the system; a second one $640$ ps long
with a $4\times 10^{-2}$ ps saving time has been used for some
intermediate frequency analysis.
The last one, with length $t_{prod}$ and saving time $t_{save}$
dependent on temperature, has been used to calculate static quantities
and the long time behavior of the system. 
\end{itemize}
All the calculations have been performed on a cluster of four $\alpha$-CPU
with a frequency of 500 MHz; every nanosecond of simulated dynamics
needed approximately 24 hours of CPU-time.

\section{RECALL OF MAIN RESULTS FROM THE MODE-COUPLING THEORY}
\label{MCT}

A great improvement to our understanding of the glassy state of matter
has come from the extension of the theoretical building
of the Mode Coupling Theory (MCT) \cite{mct1,mct2} 
developed for the {\it equilibrium}
description of the dynamics of simple, i.e. monatomic, liquid to
the study of the glassy state. Although such theory is the only
microscopic approach to the glass transition leading to many predictions
on the experimental data, it is still at the center of a strong debate 
and some questions stay open. 
Infact, even if the real range of validity of MCT
for the study of molecular liquids 
has been cleared in the last years 
(see, among others, refs.~\cite{fablatz,kamkob,singoe}),
some experimental results, like the presence of the so-called 
knee characterizing
the low-frequency behavior of the light scattering
susceptibility~\cite{nomct1,nomct2} or the presence of a cusp
in the non-ergodicity parameter~\cite{nomct3}, seem to contradict 
fundamental predictions of the idealized version of MCT.

In this section we sum up the main
predictions of the so called {\it ideal} MCT 
where it is hypothesized a complete dynamical freezing and the so called
"thermally activated hopping" processes are neglected;
such predictions will be compared with our simulation data.
In the ideal MCT the glass formation
is interpreted as a {\it dynamical transition} 
from an ergodic to a non-ergodic behavior at a 
cross-over temperature $T_c$.
MCT provides a self-consistent dynamical treatment~\cite{mct1} 
for the density correlation function of an isotropic system 
\begin{equation}
F(q,t)=\frac{1}{N}\langle\delta\rho^*_q(t)\delta\rho^*_q(0)\rangle
\end{equation}
where $N$ is the number of the particles, $\delta\rho=\rho - <\rho>$,  
$\delta\rho_q(t)=\sum_{i=1}^{N}\exp(i\bar q \cdot \bar r_{i}(t))$
and $\bar r_{i}(t)$ is the position of particle $i$ at time $t$.
MCT proposes a particular {\it ansatz} for the memory kernel
in the related integro-differential generalized Langevin equation,
such kernel is coupling non-linearly the density fluctuations
with one another. If the coupling increases upon lowering the temperature,
the resulting dynamical feedback leads to a progressive slowing 
down of the density fluctuations  until they become completely frozen 
at the critical temperature $T_{c}$. 
The ideal MCT describes the behavior as much as 
the temperature approaches $T_c$, i.e. the  
parameter $\sigma=(T_{c}-T)/T_{c}$ is small 
(however real comparisons have to be made 
for $\sigma$ not too small, in contrast to 
the case of scaling laws in phase transitions).
For temperature $T\geq T_c$, $F(q,t)$ is characterized by two 
step decays taking place at different time scales and the theory gives 
specific predictions for such different time regions. 
 
The first one, the {\it $\beta$-process} region, is centered 
around a time $\tau_{\sigma}$ which is predicted to scale like 
$\tau_{\sigma}\propto\vert T-T_{c}\vert^{1/2a}$
with $0<a<0.5$ and to be bounded in the interval
$\tau_0<<\tau_\sigma<<\tau_\alpha$ where $\tau_0$ is the time scale
of the microscopic dynamics and $\tau_\alpha$
is the structural rearrangement time scale.
In this region the {\it factorization property} holds, in the sense that
the density correlation function can be written as
\begin{equation}
F(q,t)=f(q)+h(q) \sqrt{\vert\sigma\vert} G_{\pm}(t/\tau_{\sigma})
\label{betaregion}
\end{equation}
where $f(q)$ is the {\it non-ergodicity parameter}
(i.e. {\it Debye-Waller factor} for 
collective correlators or {\it M\"ossbauer-Lamb factor} 
for single-particle correlators), $h(q)$ is an amplitude
independent of temperature and time and the $\pm$ in $G_{\pm}$
corresponds to time larger or smaller with respect to $\tau_\sigma$.
So, the time dependence of the correlation functions
is all embedded in the {\it q-independent} function  $G_{\pm}$,
namely spatial and temporal correlations result to be completely independent.
$G_{\pm}(t)$ is asymptotically expressed by two power laws,
respectively the {\it critical decay} 
and the {\it von Schweidler law}~\cite{mct1},
characterized by the {\it temperature and momentum independent} 
exponents $a$ and $b$
\begin{equation}
G_{\pm}\left(\frac{t}{\tau_{\sigma}}\right)=\left\{
\begin{array}{rr}
(t/\tau_{\sigma})^{-a} & \mbox{$\tau_0<<t<<\tau_\sigma$}\\
-(t/\tau_{\sigma})^b   &\mbox{$\tau_\sigma<<t<<\tau_\alpha$};
\end{array}
\right. 
\label{abpower}
\end{equation}
here $a$ is the same exponent of the power divergence of $\tau_{\sigma}$
at $T_{c}$ and it is related to the exponent $b$ ($0<b\leq 1$) via the
equation 
\begin{equation}
\frac{\Gamma^{2}(1-a)}{\Gamma(1-2a)}=
\frac{\Gamma^{2}(1+b)}{\Gamma(1+2b)}
\end{equation}
where $\Gamma$ is the gamma function.

The second time region is the so-called 
{\it $\alpha$-region} where the second decaying step takes place.
This region is connected to 
the collective structural relaxations 
and asymptotically the theory predicts the validity of 
the well known {\it time-temperature superposition principle};
it states that, on time scales of same order of magnitude of 
$\tau_{\alpha}$, the following scaling law holds at every temperature $T$
\begin{equation}
F(q,t)={\cal F}\left(\frac{t}{\tau_{\alpha}(T)}\right);
\label{superposition}
\end{equation}
in other words, the correlation functions of any observables
at different temperatures can be
collapsed into a master curve when the time is scaled with
$t/\tau_\alpha$. Moreover MCT predicts that this master curve  
can be fitted by a Kolrausch-William-Watts function
({\it stretched exponential})
\begin{equation}
F(q,t)\simeq f(q)\,\exp\left\{-\left(\frac{t}{\tau_\alpha}\right)
^{\beta_\alpha}\right\}
\label{stretched}
\end{equation}
The $\alpha$ time scale $\tau_\alpha$ depends on 
temperature trough a power law of the form
\begin{equation}
\tau_\alpha\propto(T-T_c)^{-\gamma}
\label{powerlaw}
\end{equation}
where the {\it q-independent} exponent $\gamma$ 
is related to the power exponents 
$a$ and $b$ of the $\beta$-region by the relation
\begin{equation}
\gamma=\frac{1}{2a}+\frac{1}{2b}.
\label{abgamma}
\end{equation}
The inverse of the diffusion constant $D^{-1}(T)$
is predicted to scale like $\tau_\alpha$ \cite{mct1} and consequently
it follows Eq.~[\ref{powerlaw}].

Up to now, all dynamical results reviewed are universal in the sense
that they are predicted to hold for the correlators 
of every observable with
non-zero overlap with density; in particular 
this is true for both the one-particle
and the collective density correlation functions.
Nevertheless important differences 
are predicted to hold for the q-dependence
in these two cases: in the former case $f(q)$ and $h(q)$ 
depend smoothly on $q$, in the latter one they oscillate 
respectively in phase and out of phase with the static
structure factor $S(q)$.
Moreover, $\beta_{\alpha}$ is predicted to be a smooth function of $q$
in the one-particle case; at variance,
it shows pronounced
oscillations in phase with $S(q)$ in the collective case.  

\section{RESULTS}
\label{results}
\subsection{THERMODYNAMICS}

In this subsection some thermodynamical time-independent results are shown 
like potential, kinetic, and total energy (see Tab.~\ref{thermotab}
and Fig.~\ref{thermo}). 
The interest of these results is clarified by the following argument:
in computer simulations dealing with the glass transition it is possible to
define a temperature often named $T_{g-sim}$ \cite{koband1} 
at which one-time quantities
show some sort of discontinuity. Such 
discontinuity, whose position depends on the thermal history of the system, 
represents the thermodynamical point at which 
the system undergoes a glass transition on the time scale of the computer 
simulation, falling out of equilibrium. It is clear from Fig.~\ref{thermo}
that no discontinuity is present, i.e. $T_{g-sim}$ of our simulations
is less then the lowest temperature studied and then we have good chance 
to have well-thermalized results.

\noindent Nevertheless, whether or not the system is in equilibrium
can be checked only {\it a-posteriori} by comparing the
total simulation time with the measured relaxation time.

From the linearity of $E(T)$ we deduce a {specific heat} 
$c(T)$ constant in the temperature range investigated
and equal to $140\pm\; J K^{-1}mol^{-1}$; 
such value must be compared with the experimental
value of $341.7\, J K^{-1}mol^{-1}$ \cite{chang}.
It is possible to explain the inconsistency between the two values
keeping in mind that our MD value is a classical (non-quantic) result
and, more important, we are neglecting many ($\sim 78$) degrees of 
freedom concerning the deformations of the phenyl rings.

\subsection{STRUCTURE}

In general the static structure of a fluid is well described by 
the {\it pair distribution function} \cite{hansen}:
\begin{equation}
g( r )=\frac{V}{N^{2}}\langle\sum_{i}\sum_{j\neq i}\delta
(r-\vert {\bar r}_{ij} \vert)\rangle.
\end{equation}
In computer simulations \cite{allen}, we can identify the distances 
$\vert {\bar r}_{ij} \vert$ with different quantities.

\noindent In Fig.~\ref{gdr} we report some $g(r)$'s at $T=300$ K
where we have considered as $\vert {\bar r}_{ij} \vert$ the distances between
the Carbon atoms belonging to different rings (A) and 
between the center of masses of rings (B) and molecules (C); 
both total (solid line) and intermolecular (dashed line)
contributions are shown  
in order to separate the internal molecular structure and the 
mean structural organization of the whole bulk sample.   
\noindent In Fig.~\ref{gdr} (B) a two peak structure is present: 
the first sharp peak is placed
at $r\simeq 0.42$ nm  corresponding to the mean distance 
between rings belonging
to the same molecule; the second one, of intermolecular origin,
is placed at $r\simeq 0.6$ nm. It is worth noting that such
distance is less than the greatest intramolecular 
C-C distance ($\simeq 0.7$ nm ). Moreover
the molecular centers of mass $g(r)$
also shows a large value on distances less than $0.7$ nm  
giving the evidence 
of a strong packing of the molecules. All these features appear to
be approximately temperature independent. 
Such packing depends strongly on the orientational internal
configuration of the molecules namely on the positions of the two side rings
with respect to the parent one; the computation 
of the probability distribution of the scalar products 
among the versors $\hat l_\xi$, $\hat m_\xi$ and $\hat n_\xi$,
introduced in sect.~\ref{molecular model}, 
is somehow instructive in this sense.

In Fig.~\ref{distrang} the distribution functions for 
the quantities $\hat l_{1}\cdot\hat n_{2,3}$, $\hat n_{2}\cdot\hat n_{3}$, 
$\hat n_{1}\cdot\hat n_{2,3}$ are shown. The first two distributions  
are practically temperature independent and give us only
informations on the correctness of the simulated geometry:
they are sharply peaked on the the correct equilibrium positions 
of about $0.71$ and $0.69$ respectively. At variance with the distribution 
of $x=\hat n_{2}\cdot\hat n_{3}$ and of $\hat n_{1}\cdot\hat n_{2,3}$
that are symmetric around $x=0$,
the distribution of
$\hat l_{1}\cdot\hat n_{2,3}$ does not present 
the symmetric peak on negative values so that 
we can argue that the auxiliary term $V_{A}$ worked correctly.
The most interesting distribution is the third one in which 
the peak intensity (the peak is correctly placed at $\alpha_{0}=0.59$)
is higher the lower the temperature, as shown in Fig.~\ref{distrnorm2}, 
indicating therefore that the correspondent degree of freedom
(the in-phase motion of $\hat n_{1}\cdot\hat n_{2,3}$) is  
more and more frozen on its equilibrium value with decreasing the temperature.

We have seen that in the isolated molecule the rotational motion
of the two side rings 
can be separated in two contributions: an out-of-phase harmonic libration and
an in-phase pivoting around the bonds which permits rings to cross from 
one equilibrium position to the other degenerate one.
It is clear from the structure of the distribution function
in proximity of $\hat n_{1}\cdot\hat n_{2,3}=0$ shown in Fig.~\ref{distrnorm}  
that the time needed for the transition 
from a minimum to the other one will be longer lowering the temperature;
moreover Fig.~\ref{distrnorm} can be considered as a restatement 
of the energy map 
shown in Fig.~\ref{map}, being the intensity of the maximum in zero 
a measure of the transition probability between the two minima. 
Such fenomenology will be clarified
in future communications where we will study the relaxation processes 
associated to the angular degrees of freedom.   

The space Fourier transform of $g(r)$ is the {\it static structure factor}.
In a poliatomic system this quantity is defined as 
\begin{equation}
S(q)\propto\frac{1}{N}\sum_{i,j}b_{i}\;b_{j}\langle e^{ i {\bar q}\cdot
({\bar r}_{i}-{\bar r}_{j})}\rangle;
\label{essek}
\end{equation}
where the coefficients $b_i$ are the  
{\it scattering lengths} in principle different for each species involved.

\noindent The $S(q)$ has been determined experimentally for OTP
by neutron scattering \cite{barbertchi,tollschowutt},
and the following main features have been observed:

\noindent {\it i}) in contrast to atomic systems its main peak 
is splitted in two sub-peaks placed around $14$ nm$^{-1}$ 
and $19$ nm$^{-1}$

\noindent {\it ii}) in the $q\rightarrow 0$ region, lowering the temperature,
a reduction of scattering intensity is
observed due to the decrease of the isothermal compressibility $\chi_{T}$
($\chi_{T}\propto S(q=0)$)

\noindent {\it iii}) increasing the density, a slight shift 
of the peak position to higher $q$ values is observed 

\noindent {\it iv}) on decreasing temperature, the height 
of the peak around $19$ nm$^{-1}$ increases 
while the intensity of the peak at $14$ nm$^{-1}$ remains nearly unaffected
except for a slight reduction mostly 
connected to the decrease in $\chi_{T}$.

In Fig.~\ref{sk_cm_rng} we show our results for the
structure factors calculated
assuming as scattering centers the
molecules and the rings centers of mass with $b_i=1$;
every point is an average on all the independent Miller indices
corresponding to the given $q$.

\noindent It is much more interesting to make a comparison among the MD results
and the experimental data obtained by neutron (Fig.~\ref{essekexp} (A))
and to test what is expected for X-Ray scattering (Fig.~\ref{essekexp} (B)).

In evaluating $S(q)$ by computer simulation 
for a comparison with neutron data,
we have to take in account the contribution due to both
Carbon and Hydrogen atoms;
$H$ atoms are not considered in our dynamics, nevertheless it is possible
to place them in fixed positions on the line extending from the center of the ring trough a carbon atom at fixed $C-H$ distance computed to be 
$d_{\;C-H}=0.107$ nm. In this case we would have to consider different 
scattering lengths for the two species, $b_H$ and $b_C$;
nevertheless they are both positive 
and about of the same magnitude so that the product
$b_{i} b_{j}$ in Eq.~(\ref{essek}) is an ineffective constant. 

In Fig.~\ref{essekexp} (A) 
the calculated $S(q)$ at $T=300\,K$ is shown and compared 
with the data of ref.~\cite{barbertchi} at $T=324$ K; in this paper
the authors show their results in terms of the {\it coherent scattering
cross section} $(d\sigma/d\Omega)$ measured in $m^2$ which is proportional
to our $S(q)$.
In order to compare the two results we renormalized
the experimental data in such a way the values of the two curves
coincide at large $q$. 
The high-q region of the calculated $S(q)$ appears to be in excellent
agreement with the experiment but no double peak structure 
is present at low momenta. In particular the MD calculated first peak presents
a small bump at about $18$ nm$^{-1}$;
this is better seen in Fig.~\ref{sk_err} where we show
the small-$q$ part of $S(q)$ calculated at $T=280\,K$ together
with the error bars estimated by means of 
the statistical fluctuation of the data.
The noise
cannot allow us to determine the correct structure of the
main peak. It is worth noting, moreover, that at this low temperature
the characteristic relaxation time is of order $1$ ns,
so that, considering a simulation run $10$ ns long, we have only
about $10$ really independent system configurations.

In order to calculate the simulated $S(q)$
as it is expected by X-ray scattering, we consider only the Carbon atoms; 
also in this case no double-peak structure is observed in the data 
but a clear prepeak appears at a q-value less then the $q$ of the 
first maximum, 
being the high-q behavior similar
to the neutron case, as shown in Fig.~\ref{essekexp} (B).

\subsection{SELF DIFFUSION COEFFICIENT}

An important quantity to consider in the study of 
the dynamics of our system at a 
microscopic level is the {\it mean squared displacement} (MSD) defined as
\begin{equation}
\langle r^{2}(t)\rangle=\frac{1}{N}\sum_{i\xi}\langle
\vert{\bar R}_{i\xi}(t)-{\bar R}_{i\xi}(0)\vert^{2}\rangle
\end{equation}
where ${\bar R}_{i\xi}(t)$ is the position of the 
center of mass of the ring $\xi$ in the molecule $i$ at time $t$;
from the MSD is possible to determine the {\it self-diffusion coefficient}
$D(T)$ via the Einstein relation 
\begin{equation}
D=\lim_{t\rightarrow \infty}\frac{1}{6t}\langle r^{2}(t)\rangle 
\label{einstein}
\end{equation}

The temperature dependence of the MSD is shown in Fig.~\ref{erre2};
each curve follows the usual {\it cage-effect} scenario. 
At small time (less than 
$0.2\; ps$) they present the $t^{2}$ behavior corresponding to the
ballistic motion; at long time the diffusive linear time dependence
of Eq.~(\ref{einstein}) is found. At intermediate times a small 
region is present where MSD stays almost constant 
and  whose duration increases with 
decreasing temperature; on these time scales molecules are trapped in cages
builded up by their neighbors, and they can only vibrate 
in these limited region, the length of the plateau being a measure
of the mean life-time of the cages.

The calculated values of the self-diffusion coefficient 
are shown in Tab.~\ref{diff_tab} and plotted in Fig~\ref{diff_fit}
(open circles) as a function of temperature,
together with the power-law temperature dependence (solid line) 
predicted by the MCT 
\begin{equation}
D^{-1}(T)\propto(T-T_{c})^{-\gamma}.
\label{power}
\end{equation}
A three parameters fit to these data has been performed 
obtaining the following values
\begin{eqnarray}
T_{c}^{(D)}  &=& 278  \pm 3 \\
\gamma^{(D)} &=& 1.8 \pm 0.1
\end{eqnarray}
In the same figure we also show the experimental data 
(full squares) \cite{calldougfal,fujgeilsill}  
that are well represented by Eq.~(\ref{power}) (dashed line) 
with the values $\gamma=2.3\pm 0.1$ and $T_{c}=292\pm 2$ K. 

It is clear from these values and from Fig.~\ref{diff_fit} that
a discrepancy is present among the 
lower temperatures diffusive behavior of
the simulated and real system respectively;
this is most likely due to the fact that we have tuned the 
value of the L-J potential parameters $\epsilon$ and $\sigma$
in order to reproduce the high-T diffusion properties of the
real system. However it is worth noting that 
it is possible to reproduce quite well  
the experimental results
on the whole investigated temperature range
shifting the molecular dynamics points at temperatures $\sim 20$ K 
above their true values. In other words we have to assume
that our actual thermodynamic point is shifted with respect
to the real one; from now on, whenever we will compare
our molecular dynamics results with the
experimental ones, our calculated points will be 
shifted of $20$ K above the measured temperature
and the competing temperatures will be indicated as ${\bar T}_c$.
On this grounds from the previous study of the
self diffusion properties of our model we obtain
${\bar T}_c^{(D)}=298\pm 3$ to be compared with the experimental
value $T_c=290\pm 5$.

A different way to determine the parameters
entering in the power law reported in Eq.~\ref{power} 
is possible, even if 
not independent from the previous one; it is based on the study of the  
{\it non-gaussian} parameter $\alpha_{2}(t)$ defined as 
\cite{odagaki,sciorgalltar}:
\begin{equation}
\alpha_{2}(t)=\frac{9}{5}\frac{\langle r^{4}(t)\rangle}
{\langle r^{2}(t)\rangle^{2}}-1
\end{equation}
where the mean square displacement $\langle r^{2}(t)\rangle$ 
and $\langle r^{4}(t)\rangle$
are, respectively,  the second and forth momenta
of the {\it  Van Hove self-correlation function}
\begin{equation}
G_{s}({\bar r},t)=\frac{1}{N}\sum_{i\xi}\langle\delta 
({\bar r}-{\bar R}_{i\xi}(t)+{\bar R}_{i\xi}(0))\rangle.
\end{equation}  
The parameter $\alpha_{2}(t)$ quantifies the degree of non-Gaussianity
of $G_{s}({\bar r},t)$ in space 
as a function of time 
and it is normalized in such a way that,
if $G_{s}({\bar r},\tilde{t}\,)$ was a
gaussian function in space at a given 
time $\tilde{t}$, we would have $\alpha_{2}(\tilde{t})=0$.
The time dependence of $\alpha_2(t)$ at all temperatures 
investigated is shown in Fig.~\ref{alpha2}.
We are not interested here in the specific time dependence of such function
but only in the fact that $t_{\alpha_{max}}$, the position of the 
maximum of $\alpha_{2}(t)$, has the power 
dependence on $T$ similar to that of
Eq.~(\ref{power})~\cite{odagaki} (see Fig.~\ref{alpha_fit}). 
A fit to this quantity performed in the same way as before gives us the values
\begin{eqnarray}
{\bar T}_{c}^{{t_\alpha}} &=&300 \pm14 \\
\gamma^{{t_\alpha}}&=&1.4 \pm0.3
\end{eqnarray}  
compatible with the values 
determined by the temperature dependence of $D$, even if 
the error bars are larger in this case.

\subsection{SINGLE PARTICLE DYNAMICS}

Comparisons of the {\it coherent} (collective) and {\it incoherent}
(self) density fluctuations dynamics data 
measured by different techniques (neutron time-of-flight
and backscattering spectroscopy, photon correlation spectroscopy, depolarized
Raman and Rayleigh-Brillouin light scattering)  
with the main predictions of MCT have been reported in literature 
with great details 
\cite{petbarpuj,kiebbardeb,bartfujleg,tollwutt,hwangshen,stefpatgla}.
In this section we will study the single particle
density fluctuation dynamics of our model and we will 
compare our results with the experimental results mainly contained in 
refs.~\cite{petbarpuj,kiebbardeb} and with the MCT
predictions.

The single particle dynamics of the model is 
embedded in the {\it incoherent 
self-intermediate scattering function} defined as
\begin{equation}
F_{s}(q,t)=\frac{1}{N}\langle\sum_{i,\xi}e^{-i{\bar q}\cdot\left[
{\bar R}_{i\xi}(t)-{\bar R}_{i\xi}(0)\right]}\rangle
\end{equation}
where, again, ${\bar R}_{i\xi}(t)$ is the position of the 
center of mass of the ring $\xi$ in the molecule $i$ at time $t$.
At every temperature considered two sets of configurations, 
produced with two different storing time as
described in section~\ref{computation}, have been used to reconstruct
the whole curve.
We considered the $T$-dependence of $F_s(q,t)$ at the two 
momentum values $q=14,19$ nm$^{-1}$\,
corresponding to the first and second peaks of the static
structure factor, averaging on values of $q$ falling 
in the interval $q\pm \Delta q$ with $\Delta q= 0.2$ nm$^{-1}$.
Finally, we spanned at $T=300$ K the whole interesting $q$-space
in the interval $q=2\div 30$ nm$^{-1}$ 
(averaging on the values of $q$ falling in the same interval
$2 \Delta q$ wide).   

In Fig.~\ref{f_self} we show $F_{s}(q,t)$ for nearly all temperatures
investigated at $q=q_{max}=14$ nm$^{-1}$; 
all the curves decay to zero i.e.
the length of all the simulations allows the fluctuations 
to become completely uncorrelated. We are in a ``good''
thermodynamical equilibrium at every 
temperature, at least on the space scales corresponding to the inverse of
$q_{max}$.

\noindent At temperatures lower then $T=330$ K the relaxation follows clearly
the predicted two step pattern: on microscopic time-scales
the correlation is quadratic in time, this time scale being
the one on which the intramolecular vibrations happen;
on intermediate time scales we observe the formation of a plateau, 
whose height is the non-ergodicity parameter $f(q)$ and whose length
in time is comparable to the one of the plateau in the MSD 
$\langle r^{2}(t)\rangle$. 
On long time scales we observe the structural relaxation in the
form of a stretched exponential.
At the highest temperatures no double pattern is visible anymore
and only a nearly exponential relaxation can be recognized.   
A stretched exponential fit (see Eq.~(\ref{stretched})) 
on the structural time scale ($\alpha$-process) 
gives us the temperature dependence
of the three free parameters $\beta_{\alpha}$,$\tau_{\alpha}$ and $f(q)$.

The parameter $\beta_{\alpha}$ (circles) is shown in Fig.~\ref{beta_err}; it
appears to be nearly $T$-independent
for temperature lower than $T=400$ K 
and its mean value $\beta_{\alpha}\simeq 0.8$ 
(dashed line) has to be compared with
the experimental value  $\beta_{\alpha}=0.6$. For temperature
in the higher region it tends toward the value $\beta_{\alpha}=1$
(errors are clearly much more greater); such behavior is due to the fact
that in this temperature region it is no longer possible
to sharply separate the long-time relaxation region from the microscopic
short-time one.
The study of the temperature dependence of the 
non-ergodicity parameter $f(q)$ in the interesting region
is not possible due to our limited temperature range which
do not permit to observe the expected low-temperature ($T<T_{g}$)
harmonic Debye-like behavior, and the onset of the anomalous decrease
of $f(q)$ with increasing $T$ for $T_{g}<T\leq T_{c}$. 
Our data suggest us $T_{c} < 283$ K and the mean value $f(q)\simeq 0.7$
(dot-dashed line)
agrees with the experimental value 
determined at $T=290$ K shown in Fig.~\ref{fit1_q}.

It is worth to test the power law 
temperature dependence Eq.~(\ref{powerlaw})
for the relaxation time $\tau_\alpha$; the calculated 
relaxation times (circles) shifted of $20$ K
with respect to the measured temperatures, as explained above,
are plotted in Fig.~\ref{tau_err}
together with the experimental (full triangles) {\it shear viscosity} 
$\eta_{s}$ data ($\eta_{s}$ is expected to be 
proportional to $\tau_\alpha$) 
of ref.~\cite{giuphd}
and the theoretical fitted curve (solid line) of parameters
\begin{eqnarray}
{\bar T}^{\tau_\alpha}_{c} &=&296 \pm7 \\
\gamma^S&=&2.0 \pm0.4
\label{fitparam}
\end{eqnarray}
to be compared with the experimental results of ref.~\cite{petbarpuj}
$T_{c}=290\pm 5$ K and $\gamma=2.55$.
These values are compatible, within the statistical error, 
with the values calculated from the diffusion data
so that we can conclude that the diffusive behavior 
and the self-dynamics of our model follow
the same {\it critical} power law with ${\bar T}_{c}\simeq 297$ K
and $\gamma \simeq 1.8$.

Also the values of $\tau_{\alpha}$ for $q=19$ nm$^{-1}$ (squares)
corresponding to the second peak of the static structure function
are reported with the theoretical curve.
A fit has been performed (dashed line) only on the prefactor keeping fixed
the values of the other two parameters in order to show
that these data also are compatible with the same power law.
The crucial observation here is that the values of the 
two parameters $T_c$ and $\gamma$ are effectively q-independent 
and they can be considered universal for our model,
as predicted by the MCT.

The relaxation time $\tau_{\alpha}$ can be also used 
to test the time-temperature
superposition principle Eq.~(\ref{superposition}). 
In Fig.~\ref{rescaling}
the curves are shown in function of the rescaled time $t/\tau_{\alpha}$
and it is clearly seen that all the curves tend to collapse on the same 
master curve as predicted by the theory.

We now quantify the q-dependence of the self
dynamics long-time behavior of the system at $T=300$ K.
In Fig.~\ref{f_self_q} are reported the curves $F_{s}(q,t)$ 
for values of $q=2 n$ nm$^{-1}$ with $n=3,\ldots, 15$;
the choice of the temperature value $T=300$ K has been due to the need
of ``well-thermalized'' results in a large range of q.
Also in these data is well defined the two step behavior
and we can calculate the long time 
stretched exponential fit parameters;
the resulting values are shown in Fig.~\ref{fit1_q} and Fig.~\ref{fit2_q}.
In Fig.~\ref{fit1_q} the values of $\beta_{\alpha}$ (left side)
and $f(q)$ (right side) are shown. 

$\beta_{\alpha}$ (open circles) appears to be
a smooth function of $q$ and it tends, for large
values of $q$, to the experimental (full circles)
evaluated value of $0.6$. Such behavior is quite general
(see, for instance, ref.~\cite{sciorgalltar})
and can be easily explained by the following argument~\cite{sciorgalltar}:
for large values of $q$, corresponding to length scales
of the same order of magnitude of the cages dimension,
the dynamics becomes slower and slower approaching 
the cage dynamics described trough the von Schweidler exponent $b$.
At variance, in the opposite limit of small $q$,
we consider a diffusive dynamics on large distances;
at such length scales the decay of the self-density
fluctuations is of the usual purely exponential form 
\mbox{$\exp{(-D\,q^2\,t)}$}
corresponding to $\beta_\alpha =1$ (see Fig.~\ref{gauss_approx}). 
At this stage, however,
we have not a reasonable explanation of the disagreement
with the experimental data.

In the right side of Fig.~\ref{fit1_q} 
the q-dependence of the non-ergodicity parameter $f(q)$ is also shown
as calculated from the short time limit of the $\alpha$-process (squares),
from the long-time part (triangles) of the $\beta$-region (as
we will see below) and from the experimental data 
(full squares)~\cite{petbarpuj};
it seems clear a good
agreement between our values 
and the experimental results. 
The data appear to be monotonic decreasing as increasing $q$
and this dependence is expected to be approximately gaussian;
in Fig.~\ref{fit1_q} a gaussian fit (solid line) in the form
$\exp(-q^{2}/2\sigma^{2})$ with $\sigma\simeq 19$ nm$^{-1}$
to molecular dynamics $\alpha$-region data  is also shown.
It is clear that the $q$-range considered 
here is too limited to really decide
on the validity of this functional form (a linear approximation
would work well too), a good estimate of the error bars
lacking in this case.

In Fig.~\ref{fit2_q} the $q$-dependence of $\tau^{-1}_{\alpha}$ 
is shown (circles) together with the experimental 
data~\cite{petbarpuj}(full circles).
Molecular dynamics points
have been rescaled by a factor $\tau^{MD}_{\alpha}(T=280 K)/
\tau^{MD}_{\alpha}(T=300 K)\simeq 6.5$ to take into account 
the fact, as discussed above, that our system temperature is $20$ K
higher the real one. The correct square law behavior at low-$q$
$\tau^{-1}_\alpha(q)\simeq 6 D q^2$ (see Eq.~\ref{gaussapprox})
is also shown as a solid line; here $D$ is the self diffusion coefficient
and $6D=20.4\times10^{-5}$ nm$^2$/ps.

Finally, in Fig.~\ref{tau_D}, two products of some calculated quantity, 
expected to be constant, are shown; on the left side the statement
$D^{-1}(T)\propto\tau_{\alpha}(T)$ is proven for $q=q_{max}$ (good at
highest temperature). On the right side we show 
the product $\tau_{\alpha}\,q^{2}\,D$
evaluated at $T=300$ K which is nearly constant 
up to $q\simeq 18$ as expected, being this value approximately
the crossover point among the correct quadratic behavior 
and the asymptotically linear regime as we found above.

Let us now probe the MCT conclusions about the $\beta$-{\it region}
which is predicted to follow the power laws of Eq.~(\ref{abpower}):
we show here the results at fixed temperature $T=300$ K for values 
of $q=6+n$ nm$^{-1}$ with $n=0,\ldots,17$.

We fit all the curves by two power functions on 
the time ranges $\theta_1$ and $\theta_2$:
\begin{equation}
F_s(q,t)=\left\{
\begin{array}{cc}
f_q + c_1\, t^{-a} & \mbox{$t\in \theta_1$}\\
f_q - c_2\, t^{b}  & \mbox{$t\in \theta_2$}
\end{array}
\right. 
\label{abfit}
\end{equation}
where $\theta_1 =[0.15:2]$ (ps) and $\theta_2 =[3:20]$ (ps); 
some selected fits are shown in Fig.~\ref{beta_resc_q} 
and they seems to work quite well.

Some observations are needed on the following analysis:
a great uncertainty stems from the choice of
the fit range (i.e. from the choice of $\theta_1$ and $\theta_2$)
due to the consideration of a cross-over region
between two process not sharply separated in time;
moreover it cannot be excluded an analogous problem
between the microscopic region and the critical decay region
characterized by the exponent $a$.
Such difficulty implies a great uncertainty on the determination 
of $f(q)$ which is supposed to be considered as the long-time limit of
the $\beta$-process and the short-time behavior of the $\alpha$-relaxation. 
At this point it is clear that, lacking a careful error analysis
on the data points, such fits can only state that a parameterization 
of the data in the form of Eq.~(\ref{abfit}) is possible
which is consistent with the theory predictions \cite{kiebbardeb},
the current values of the determined parameters being 
considered only from a qualitative point of view.

Nevertheless the values of such fitting parameters,
shown in Fig.~\ref{beta_fit_q} and Fig.~\ref{h_c1_c2},
appear to be in good agreement with the values obtained
from the experimental data.
In Fig.~\ref{beta_fit_q} are shown (left) the values of the
power exponents $a$ (squares) and $b$ (circles) and of the exponent $\gamma$
(triangles)
calculated by means of Eq.~(\ref{abgamma}) (right side); 
the mean values are $0.3$, $0.5$, $2.6$ respectively, to be compared with
the experimental determined values 
$a\simeq 0.31$ (dot-dashed line), 
$b\simeq 0.52$ (dashed line), $\gamma\simeq 2.55$ (solid line) 
\cite{petbarpuj}.

It is clear from these results that one of the main prediction
of MCT, namely the $q$-independence of $a$ and $b$ is verified
in the limit of the error fluctuations.
Moreover it is important to note that the parameter $\gamma$, given by
Eq.~(\ref{abgamma}), remains constant, as expected,
being its mean value $2.6$ clearly compatible 
with the value $2.55$ determined from experimental data \cite{petbarpuj};
nevertheless this value overestimates the value 
$\gamma=2.0\pm 0.4$ at $q=14$ nm$^{-1}$
previously determined by the fit to the $\alpha$-region.
Furthermore $b(q)$ is always less then the determined value of 
the large-$q$ value of the stretching parameter $\beta_\alpha=0.6$
(see Fig.~\ref{fit1_q})
verifying an other MCT prediction, namely $0<b<\beta_\alpha$.

From Eqs.~(\ref{betaregion})~(\ref{abpower})~(\ref{abfit})
the two parameters $c_1(q)$ and $c_{2}(q)$ result to be proportional to $h(q)$,
the proportionality constants being dependent on $\sigma$, $\tau_\sigma$,
$a$ and $b$. From $T_c\simeq 280$ K we have $\sqrt{\sigma}\simeq 0.3$
while we choose as a good estimate of $\tau_\sigma$ the intersection
point of the two power laws of Fig.~\ref{beta_resc_q}, obtaining
$\tau_\sigma\simeq 2$ ps; if we put $a=0.31$ and $b=0.52$
we finally obtain the factors $2.7$ and $2.3$ for $c_1$ and $c_2$
respectively.

Unfortunately these value are not able to correctly rescale
our data on the experimental results, the correct values
being $0.7$ and $4$ as shown in Fig.~\ref{h_c1_c2};
this result could be expected considering the great uncertainties
on the parameter values used for the estimate.
Nevertheless, simulated data 
agree quite well with the experimental points
at low-$q$ presenting a strong bending toward a constant value 
in the region $q>16$ nm$^{-1}$.

To complete the picture of the self-motion in our model,
we test the validity of the {\it gaussian approximation}
to $F_s(q,t)$
in the limit of small momentum $q$. The first order term of the
expansion of $F_s(q,t)$ in powers of $q^2$ gives \cite{hansen}
\begin{equation}
F_{s}(q,t)\simeq \exp\left\{-\frac{q^{2}}{6}\langle r^{2}(t)\rangle\right\}.
\label{gaussapprox}
\end{equation}
In Fig.~\ref{gauss_approx} some curves at $T=330$ K and
different values of $q$ are shown together with the corresponding
approximations; such approximation seems to work quite good
and it becomes worse on increasing $q$ as expected.

At the end, we show in Fig.~\ref{tau_global} all the time scales
related to centers of mass motion investigated up to now
as a function of temperature.
Full circles and squares indicate respectively the 
experimental structural relaxation time $\tau^{V-F}_{EXP}(T)$,
following the Vogel-Fulcher law, and the experimental inverse 
self-diffusion coefficient $D^{-1}_{EXP}(T)$ 
multiplied by a factor $5\times 10^{-5}$ cm$^2$ in order to superimpose
to $\tau^{V-F}_{EXP}(T)$. The open symbols are used
to represent the molecular dynamics results:
$\tau^s_{MD}$ (diamonds) is the relaxation time of the one-particle dynamics
at $q=14$ nm$^{-1}$ multiplied by a scale factor $1.5$ and 
$D^{-1}_{MD}(T)$ (triangles up) is the inverse of the diffusion coefficient
rescaled by the same factor $5\times 10^{-5}$ cm$^2$ 
we used for the experimental points.
All the molecular dynamics points have been shifted of $20$ K 
with respect to the measured temperatures;
it is quite clear that both experimental and molecular dynamics data points 
collapse on a well defined master curve. 
Our model is, at least, a good model for centers of mass
dynamics of OTP.

\section{CONCLUSIONS}
\label{conclusions}

In this paper we have introduced a new interaction potential model
capable to describe 
the intramolecular dynamics of the fragile glass former OTP;
such model appears to be much more efficient with respect to
the ones introduced so far in the sense that it represents
a much more better compromise between the resulting computing
needing and its capability to mimic all the complexity
of the dynamical behavior of the real system.

It takes into account not only the translational 
and rotational dynamics of the molecules as a whole 
but also the stretching along the molecular bonds,
the tilt of the bonds, the rotations of the side rings 
with respect to the parent ring. It is tuned
in such a way to reproduce the isolated molecule vibrational
spectrum.  In this way, most likely,
we have introduced the degrees of freedom whose interplay causes 
the complex dynamical behavior of the real system.

We have, then, presented the results of molecular
dynamics computer simulations of such model;
we mainly studied the 
static structure of a bulk sample, the 
self diffusion properties and the self part of the
density-density correlation functions.

The static structure factor simulated in such a way is
compared with the experimental measures and shows
a good agreement with the neutron scattering data except for 
the very low momenta behavior (due, probably, to the finite size
of our system). Moreover we have no evidence
of the splitting of the main peak 
in two subpeak placed at $q=14, 19$ nm$^{-1}$,
only the first one being clearly visible. This luck may be due
mainly to the temperature range investigated (the intensity 
of the second sub-peak increases with lowering temperature).

The self-diffusion properties of the system have been investigated
trough the mean squared displacement and the self diffusion
coefficient temperature behavior; comparisons with experimental 
self-diffusion data give a very good agreement, showing the evidence
of compatible critical dynamics behavior in approaching
the instability temperature $T_c$ of MCT which is here find
to be $T_c=278\pm 3$ to be compared with the experimental value
$T_c=290\pm 5$ determined by a MCT analysis of the dynamics 
of the density fluctuations, a discrepancy that is most to likely ascribed
to the intermolecular LJ potential parameters $(\sigma,\epsilon)$
that have been tuned in the temperature region close to $T=300$ K. 
Moreover we considered 
the critical temperature dependence of the so-called non-gaussianity parameter
$\alpha_2(t)$ obtaining compatible values for the power law
parameters. 

The self dynamics of the density fluctuations has
been studied in great detail on the whole accessible time window,
spanning the range from time scale of the order of few fs
to times of order of some ns; moreover its dependence on temperature
and momenta has been investigated.
All the correlation curves calculated show the
typical two step behavior predicted by MCT,
the first one on short time being associated to the 
so-called ``microscopic'' processes,
i.e. the vibrational motion of molecules in the
cage built up by their neighbors; the second one being
associated to the $\alpha$-process which controls the
structural rearrangements of molecules on long time scale.
 
The critical dynamics on $\alpha$ time scale approaching
the correspondent $T_c$
is in good agreement with experimental findings; indeed
the estimate values for our model 
of the exponent $\gamma=2.0\pm 0.4$
must be compared with the experimental value $\gamma = 2.5$.

The q-dependence at $T=300$ K in the momentum range $q=6-30$ nm$^{-1}$
has been analyzed in terms of a stretched exponential fit; the values of the
determined parameters are in good agreement with the ones 
calculated by fitting the experimental points and with the MCT expected
behavior.

To summarize, in the present work we have shown
that our model for OTP fluid is mimicking rather well
the center of mass dynamical features 
of the real system,
giving results in most cases fully compatible with the experimental
findings. It is clear that its skill to help us in the understanding
of the most exotic dynamic features of the real systems
and, in particular, of the relevance of the internal degrees of freedom
on the translational dynamics,
has not been cleared in this paper; problems like the 
collective density  fluctuation behavior, the rotational dynamics,
the origin of the 
unusual fast relaxational dynamics and many others, 
will be addressed in future works~\cite{phdste}.

\begin{center}
{\bf ACKNOWLEDGEMENTS}
\end{center}

The authors wish to thank W.~G\"{o}tze for 
a critical reading of the manuscript,
L.~Bernardini and G.~Giuliani of the ``Parco Tecnologico d'Abruzzo'' 
for technical support, 
G.~Monaco and F.~Sciortino for some useful discussions.

\newpage

\begin{center}
{\bf TABLE CAPTIONS}
\end{center} 

\begin{enumerate}

\item {\footnotesize Values of the internal potential coefficients
determined by a least square minimization of the error function,
Eq.~(\ref{chiquadro}). The coefficients $c_k$
are the force constants associated to the different intramolecular
potential terms; the $b_k$, derived from the $\{c_k\}$ 
via the Eq.~(\ref{bval}),
describe the form of the in-phase
rotational potential according to Eq.~(\ref{vr1}).
}
\label{coeff}

\item {\footnotesize Free molecule vibrational frequencies:
in the second column are reported the values $\omega_{HF}$
determined by a Harthree-Fock calculation of the ground state
of the isolated molecule (we consider only the $12$ lowest eigenvalues);
in the third column are shown 
the frequencies $\omega_{MD}$ corresponding to the different peaks  
of the spectrum determined from the atomic 
velocity autocorrelation function calculated
by means of a preliminary molecular dynamics simulation 
of the isolated molecule ($\epsilon=0$) at $T=1$ K.
}
\label{frequencies}

\item {\footnotesize Simulation runs details:
at each temperature $T$, the system is coupled for a time $t_{resc}$
to a stochastic heat-bath, then it evolvs freely, at constant energy,
for a time $t_{term}$. At the end of this process, we consider 
the final configuration to be in a good equilibrium state
and we start a trajectory $t_{prod}$ ps long, saving a system configuration
every $t_{save}$ ps.
}
\label{protocol}

\item {\footnotesize Some thermodynamical results: temperatures 
$T$ effectively measured, total energy $E_{tot}$, 
total potential energy $V_{tot}$ and kinetic energy $T$.}
\label{thermotab}

\item {\footnotesize Temperature dependence of the molecular dynamics 
self-diffusion coefficient $D$.}
\label{diff_tab}

\end{enumerate}
\newpage

\begin{center}
{\bf FIGURE CAPTIONS}
\end{center}

\begin{enumerate}

\item {\footnotesize Molecular structure of OTP ($C_{18}H_{14}$);
it is constituted by three phenyl rings, the two side rings
being attached to the parent (i.e. central) ring by covalent bonds.}
\label{OTP}

\item {\footnotesize Model geometry: each phenyl ring 
is represented by a rigid hexagon
of side $L_{a}=0.139$ nm  and the equilibrium bond 
lenght is $L_{b}=0.150$ nm. $C_1$, $C_2$, $C_3$ represent the 
origins of the reference frames fixed with the rings;
$\bar{u}_{2}$, $\bar{u}_{3}$, $\bar{u}_{1(2)}$, $\bar{u}_{1(3)}$
are the vectors parallel to the ring bonds; $\hat{l}_1$ and $\hat{m}_1$
are two versors identifying the parent ring plane; $\bar P_{2}$,
$\bar P_{3}$, $\bar P_{1(2)}$, $\bar P_{1(3)}$ are the positions of 
the Carbon atoms bonding together the rings; $\bar S_{2}$,
$\bar S_{3}$, $\bar S_{1(2)}$, $\bar S_{1(3)}$ are four interaction
sites introduced to force the rings towards 
the co-planar equilibrium condition; the symbols
$\bar P_{A}$ and $\bar P_{B}$ have been introduced
to identify the angles
$\bar P_2-\bar P_{1(2)}-\bar P_{A}$ and $\bar P_3-\bar P_{1(3)}-\bar P_{B}$.} 
\label{geometry}

\item {\footnotesize Rotational energy: each point has been determined by 
an {\it ab-initio} Hartree-Fock calculation of the potential energy surface
as a function of the rotational angles $\phi_1$ and $\phi_2$
(expressed in degrees)
with all the other degrees of freedom fixed to their equilibrium values;
the energy is expressed in Hartree units (1 Hartree=27.2eV). 
We quote by A the isolines ranging from $0.0505$ to $0.0545$ mHartree
with an increment $\Delta_1=0.005$ mHartree, by B the ones
from $0.055$ to $0.095$ mHartree with $\Delta_2=0.05$ mHartree
and by C those
from $0.5$ to $1$ mHartree with $\Delta_3=0.5$ mHartree.
This figure has to be considered only from a semi-quantitative point of view
as explained in the text. 
}
\label{map}

\item {\footnotesize Temperature dependence of the energies 
tabulated in Tab.~\ref{thermotab}, $E_{tot}$ (circles)
$V_{tot}$ (squares) and $T$ (triangles up)
together with the internal $V_{intra}$ (triangles down) and the intermolecular
Lennard-Jones $V_{inter}$ (diamonds) contributions to $V_{tot}$.}
\label{thermo}

\item {\footnotesize Pair static distribution functions at $T=300$ K 
calculated on atoms (A), ring centers of mass (B) and
molecular centers of mass (C); full lines represent the total
contribution of both intramolecular and intermolecular distances,
dashed lines only the intermolecular contribution.
}
\label{gdr}

\item {\footnotesize Static distribution functions of the scalar products 
$\hat n_{2}\cdot\hat n_{3}$ (triangles),
$\hat l_{1}\cdot\hat n_{2,3}$ (circles), 
and $\hat n_{1}\cdot\hat n_{2,3}$ (squares)
evaluated at $T=280$ K.
}
\label{distrang}

\item {\footnotesize 
Intensity of the peaks of the static
distribution function of $\hat n_{1}\cdot\hat n_{2,3}$
at $T=440, 390, 330, 280$ K 
(from bottom to top). A blow-up of this figure around $x=0$
is shown in Fig.~\ref{distrnorm}.}
\label{distrnorm2}

\item {\footnotesize Temperature dependence of the static distribution 
function of $\hat n_{1}\cdot\hat n_{2,3}$ near the saddle point position 
$\hat n_{1}\cdot\hat n_{2,3}=0$ at 
$T=440, 420, 390, 370, 350, 330, 320, 300, 280$ K 
(from top to bottom).}
\label{distrnorm}

\item {\footnotesize Static structure factors at $T=300\; K$
calculated on on the molecular (solid line) 
and ring centers of mass (dashed line); each $q$ point is the average
over all the independent Miller indeces corresponding to it.}
\label{sk_cm_rng}

\item {\footnotesize {\it Top}: Comparison among molecular dynamics structure factor
(solid line) calculated taking in account both Carbon and Hydrogen atoms
as scattering centers
and experimental structure factor (dashed line) measured by
neutron scattering (from~\cite{barbertchi}).
{\it Bottom}: Molecular dynamics structure factor (solid line)
calculated taking in account only Carbon atoms;
this should be the correct result to be compared with
experimental structure factor measured by
X-ray scattering.}
\label{essekexp}

\item {\footnotesize Enlargement of the low-momenta region of $S(q)$ 
calculated at $T=280$ K: the error bars are estimated 
from the fluctuation of the single configuration $S(q)$'s.}
\label{sk_err}

\item {\footnotesize Temperature dependence of the mean square displacement
$\langle r^2(t)\rangle$ calculated on ring centers of mass
at all temperature investigated except $T= 440, 420$ K 
(higher temperature on top). Inset: linear scale plot 
of the mean square displacement at some
selected temperatures (open symbols) together with
the long time linear behavior (dashed lines). 
}
\label{erre2}

\item {\footnotesize Temperature dependence of molecular dynamics (open circles)
and experimental (full triangles) diffusion coefficients together with the
power law fits in the form of Eq.~\ref{power} 
(solid and dashed lines respectively);
we show also the MD data shifted of $20$ K (open squares)
as explained in the text.}
\label{diff_fit}

\item {\footnotesize Time dependence of the non-gaussian parameter 
$\alpha_{2}(t)$ for all temperature investigated 
(lower temperatures on top).}
\label{alpha2}

\item {\footnotesize Power law fit of the temperature dependence of 
the position $t_{max}$ of the maximum of the non-gaussian parameter. 
}
\label{alpha_fit}

\item {\footnotesize Temperature dependence of $F_{s}(q,t)$ 
calculated at $q=14$ nm$^{-1}$
for all temperatures investigated except $T=410, 430$ K
(lower temperatures on top).}
\label{f_self}

\item {\footnotesize Temperature dependence of the stretching 
parameter $\beta_{\alpha}$ (circles) 
and of the non-ergodicity parameter $f_q$ (squares);
the horizontal lines indicate the mean values of $\beta_\alpha$ (dashed line)
and $f_q$ (dot-dashed line).}
\label{beta_err}

\item {\footnotesize Temperature dependence of $\tau_{\alpha}$
at $q=14, 19$ nm$^{-1}$ (circles and squares respectively) 
together with the power law fits with ${\bar T}^{\tau_\alpha}_{c}=296$ K and
$\gamma^{\tau_\alpha}=2.0$ (solid and dashed lines respectively); 
also the experimental shear viscosity 
$\eta_s\propto \tau_\alpha$ 
data (full triangles) are 
reported (see~\cite{giuphd} and reference therein)
multiplied by a factor $1.5$ ps/Poise.
Molecular dynamics results have been shifted of $20$ K
with respect to the measured temperatures, as explained in text.}
\label{tau_err}

\item {\footnotesize $F_s(q,t)$ at $q=14$ nm$^{-1}$ rescaled to $t/\tau_{\alpha}$;
all the curves verify the time-temperature superposition principle.}
\label{rescaling}

\item {\footnotesize Q-dependence of $F_{s}(q,t)$ at $T=300$ K
for $q=2n$ nm$^{-1}$ with $n=3,\ldots,15$ (from top to bottom).}
\label{f_self_q}

\item {\footnotesize Q-dependence of the stretching and non-ergodicity parameters:
{\it Left}: molecular dynamics (circles) and
experimental (filled circles, from \cite{petbarpuj}) values
of the coefficient $\beta_\alpha$ as determined by the KWW fits;
{\it Right}:
experimental values (filled squares) of 
the non-ergodicity parameter~\cite{petbarpuj} together with
the molecular dynamics results as determined by a MCT analysis
of both $\alpha$ (squares) and $\beta$ (triangles) regions 
and the gaussian fit (solid line) to $\alpha$-region results 
with $\sigma^2=365$ nm$^{-2}$. 
}
\label{fit1_q}

\item {\footnotesize Q-dependence of the inverse relaxation time $\tau^{-1}_{\alpha}$;
molecular dynamics data (open circles)
have been multiplied by a factor $6.5$ in
order to overlap the 
experimental~\cite{petbarpuj} data (full circles),
as explained in the text.
The solid line is the
correct small-$q$ behavior $\tau^{-1}_\alpha(q)\simeq 6 D q^2$
where 
$6D=20.4\times10^{-5}$ nm$^2$/ps.
}
\label{fit2_q}

\item {\footnotesize {\it Left}: Temperature dependence of the product 
$\tau_{\alpha}\,D$ at $q=1.4$ nm$^{-1}$;{\it Right}:
q-dependence at $T=300$ K of $\tau_{\alpha}\,D\,q^{2}$.
These quantity are expected to be constant.}
\label{tau_D}

\item {\footnotesize Power laws in the $\beta$-region
for $q=8,10,12,14$ nm$^{-1}$;
critical decay of exponent $a$ (dot-dashed lines) and von Scweidler
law of exponent $b$ (dashed lines) have been reported with
simulation points.
}
\label{beta_resc_q}

\item {\footnotesize Q-dependence of the power laws exponents:
{\it Left}: the experimental values (filled circles) for $b(q)$
are plotted together with the molecular dynamics values for
$b(q)$ (open circles) and $a(q)$ (open squares); the mean values 
for our results are shown by the dashed and dott-dashed line
respectively. {\it Right}: our values for the q-dependence of the exponent
$\gamma$ determined by the relation $\gamma=1/2a + 1/2b$.}
\label{beta_fit_q}

\item {\footnotesize Q-dependence of the experimental (full circles) 
coefficient $h$ of Eq.~\ref{betaregion} and of our fitting parameters
$c_1$ (triangles) $c_2$ (squares); note that our results have been 
rescaled by a factor $0.7$ and $5$ respectively in order to
superimpose to experimental data.}
\label{h_c1_c2}

\item {\footnotesize Gaussian approximation to $F_{s}(q,t)$ at $T=330$ K
for $q=2,3,4,6,8,1,1.2$ nm$^{-1}$ (from top to bottom).}
\label{gauss_approx}

\item {\footnotesize Master plot of the temperature dependence of all the 
centers of mass time scales
discussed: we have used full symbols for experimental results
and open symbols for molecular dynamics data. Molecular dynamics points
collapse exactly on the master curve identified by the experimental data
if they are multiplied by a scale factor 
(taking into account the momentum dependencies
of the relaxation time $\tau^s_{MD}$
and the correct dimensionality of the
diffusion coefficients) and the corresponding
temperatures are shifted
of $20$ K above the measured ones, as discussed in the text. 
In particular $\tau_{MD}^s$ (open diamonds) is multiplied by a factor $1.5$,
the self diffusion coefficients $D^{-1}_{EXP}$ (full squares)
and $D^{-1}_{MD}$ (open triangles) are rescaled by a factor
$5\times 10^{-5}$ cm$^2$.}
\label{tau_global}

\end{enumerate}

\newpage

\begin{center}
{\bf TABLES}
\end{center}

\centering
{\bf Tab.~1}

\begin{table}
\centering
\begin{tabular}{||l|rl||}
$c_{1}$ & $ 1.36 \cdot 10^{-16}$ & $J / nm^{2}$\\ \hline
$c_{2}$ & $ 6.06 \cdot 10^{-17}$ & $J / nm^{2}$\\ \hline
$c_{3}$ & $ 2.44 \cdot 10^{-19}$ & $J         $\\ \hline
$c_{4}$ & $ 2.65 \cdot 10^{-17}$ & $J / nm^{2}$\\ \hline
$c_{5}$ & $ 4.23 \cdot 10^{-19}$ & $J         $\\ \hline
$c_{6}$ & $ 7.01 \cdot 10^{-19}$ & $J         $\\ \hline\hline
$b_{1}$ & $ 3.62 \cdot 10^{-20}$ & $J         $\\ \hline
$b_{2}$ & $-4.11 \cdot 10^{-19}$ & $J         $\\ \hline
$b_{3}$ & $ 6.92 \cdot 10^{-19}$ & $J         $\\ 
\end{tabular}
\end{table}

\centering
{\bf Tab.~2}

\begin{table}
\centering
\begin{tabular}{||l|r|r||}
     & $\omega_{HF}(cm^{-1})$ & $\omega_{MD}(cm^{-1})$ \\ \hline
$1$  & $49. $ & $55.  $\\ \hline
$2$  & $54. $ & $85.  $\\ \hline
$3$  & $63. $ & $99.  $\\ \hline
$4$  & $97. $ & $148. $\\ \hline
$5$  & $114.$ & $165. $\\ \hline
$6$  & $140.$ & $172. $\\ \hline
$7$  & $256.$ & $273. $\\ \hline
$8$  & $271.$ & $302. $\\ \hline
$9$  & $291.$ & $306. $\\ \hline
$10$ & $358.$ & $376. $\\ \hline
$11$ & $386.$ & $409. $\\ \hline
$12$ & $428.$ & $436. $\\ 
\end{tabular}
\end{table}

\centering
{\bf Tab.~3}

\begin{table}
\centering
\begin{tabular}{||l|r|r|r|r||}
$T (K)$ & $t_{resc}(ps)$ & $t_{term}(ps)$ & $t_{prod}(ps)$ & $t_{save}(ps)$\\ \hline
$443$ & $100$ & $1000$ & $1000 $ & $0.1$ \\ \hline
$433$ & $100$ & $1000$ & $1000 $ & $0.1$ \\ \hline
$420$ & $100$ & $1000$ & $1000 $ & $0.1$ \\ \hline
$410$ & $100$ & $1000$ & $1000 $ & $0.1$ \\ \hline
$389$ & $200$ & $1400$ & $2000 $ & $0.1$ \\ \hline
$372$ & $200$ & $1400$ & $2000 $ & $0.1$ \\ \hline
$351$ & $200$ & $1400$ & $2000 $ & $0.1$ \\ \hline
$331$ & $500$ & $2000$ & $4000 $ & $0.2$ \\ \hline
$321$ & $500$ & $3000$ & $6000 $ & $0.3$ \\ \hline
$313$ & $500$ & $3000$ & $6000 $ & $0.3$ \\ \hline
$300$ & $500$ & $3000$ & $10000$ & $0.5$ \\ \hline
$294$ & $500$ & $5000$ & $10000$ & $0.5$ \\ \hline
$283$ & $500$ & $5000$ & $10000$ & $0.5$ \\ 
\end{tabular}
\end{table}

\centering
{\bf Tab.~4}

\begin{table}
\centering
\begin{tabular}{||l|r|r|r||}
$T (K)$ & $E_{tot}(kJ/mol)$ & $V_{tot}(kJ/mol)$ & $T(kJ/mol)$\\\hline
$443$ & $30.00 $ & $-3.15 \pm0.72$ & $33.12 \pm1.29$ \\ \hline
$433$ & $28.59 $ & $-3.78 \pm0.69$ & $32.37 \pm1.26$ \\ \hline
$420$ & $26.88 $ & $-4.59 \pm0.69$ & $31.47 \pm1.26$ \\ \hline 
$410$ & $25.47 $ & $-5.25 \pm0.69$ & $30.72 \pm1.20$ \\ \hline
$389$ & $22.44 $ & $-6.69 \pm0.63$ & $29.13 \pm1.14$ \\ \hline
$372$ & $19.98 $ & $-7.83 \pm0.60$ & $27.81 \pm1.08$ \\ \hline
$351$ & $17.10 $ & $-9.15 \pm0.57$ & $26.28 \pm1.02$ \\ \hline
$331$ & $14.34 $ & $-10.47\pm0.54$ & $24.81 \pm0.96$ \\ \hline
$321$ & $12.75 $ & $-11.22\pm0.54$ & $24.00 \pm0.96$ \\ \hline
$313$ & $11.73 $ & $-11.67\pm0.51$ & $23.43 \pm0.93$ \\ \hline
$300$ & $9.96  $ & $-12.51\pm0.51$ & $22.47 \pm0.90$ \\ \hline
$294$ & $9.15  $ & $-12.90\pm0.48$ & $21.99 \pm0.90$ \\ \hline
$283$ & $7.59  $ & $-13.59\pm0.48$ & $21.18 \pm0.84$ \\ 
\end{tabular}
\end{table}

\centering
{\bf Tab.~5}

\begin{table}
\centering
\begin{tabular}{||l|l||}
$T (K)$ & $10^{7}\times D(cm^{2}/s)$          \\ \hline
$443$ & $106.2$  \\ \hline
$433$ & $89.3 $  \\ \hline
$420$ & $81.3 $  \\ \hline
$410$ & $66.6 $  \\ \hline
$389$ & $49.3 $  \\ \hline
$372$ & $37.3 $  \\ \hline
$351$ & $27.3 $  \\ \hline
$331$ & $12.5 $  \\ \hline
$321$ & $9.4  $  \\ \hline
$313$ & $6.3  $  \\ \hline
$300$ & $3.4  $  \\ \hline
$294$ & $0.8  $  \\ \hline
$283$ & $0.6  $  \\ 
\end{tabular}
\end{table}

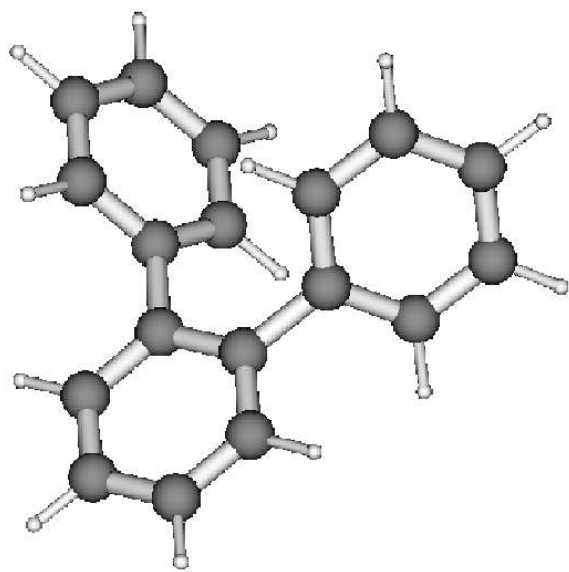
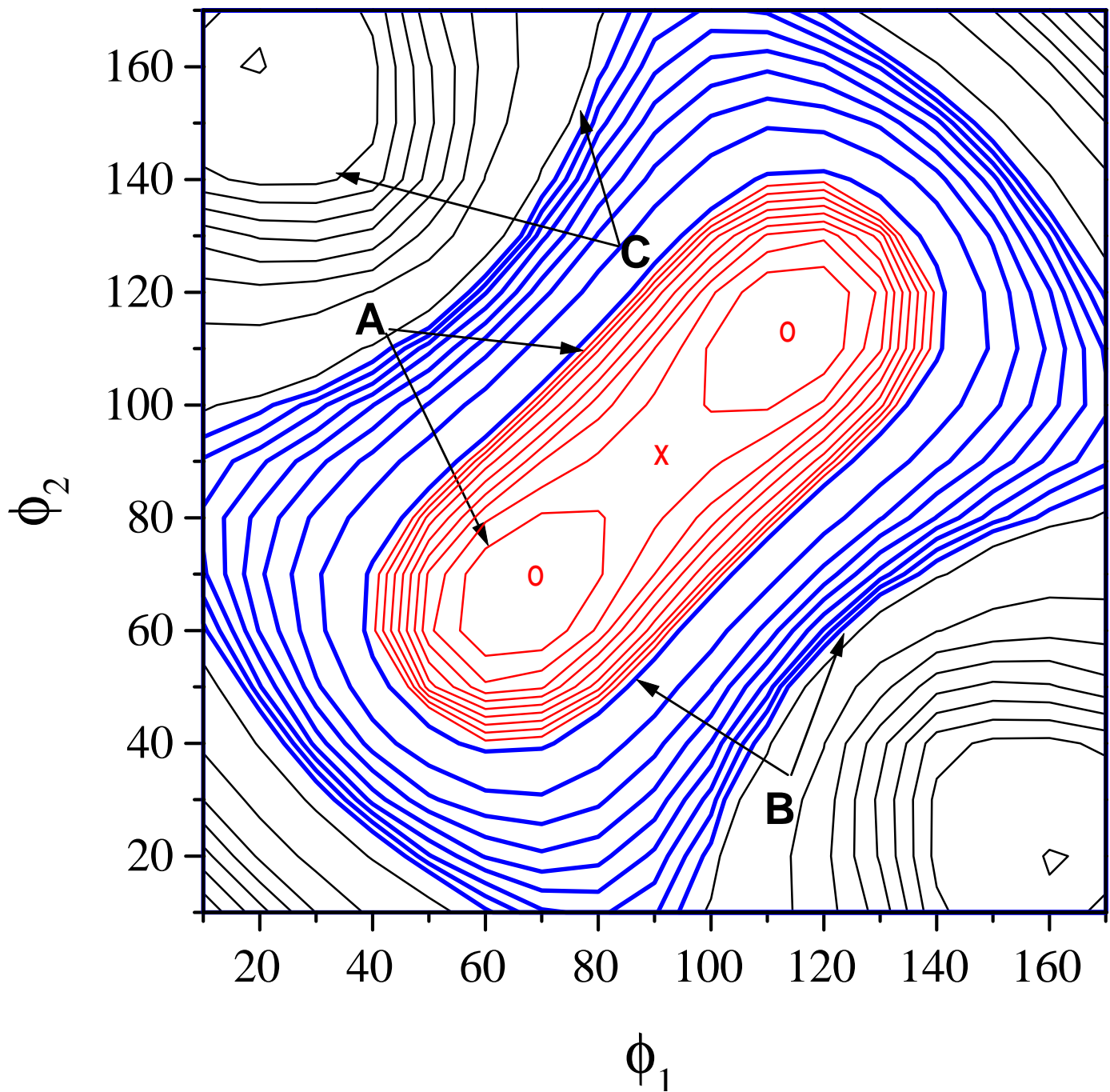
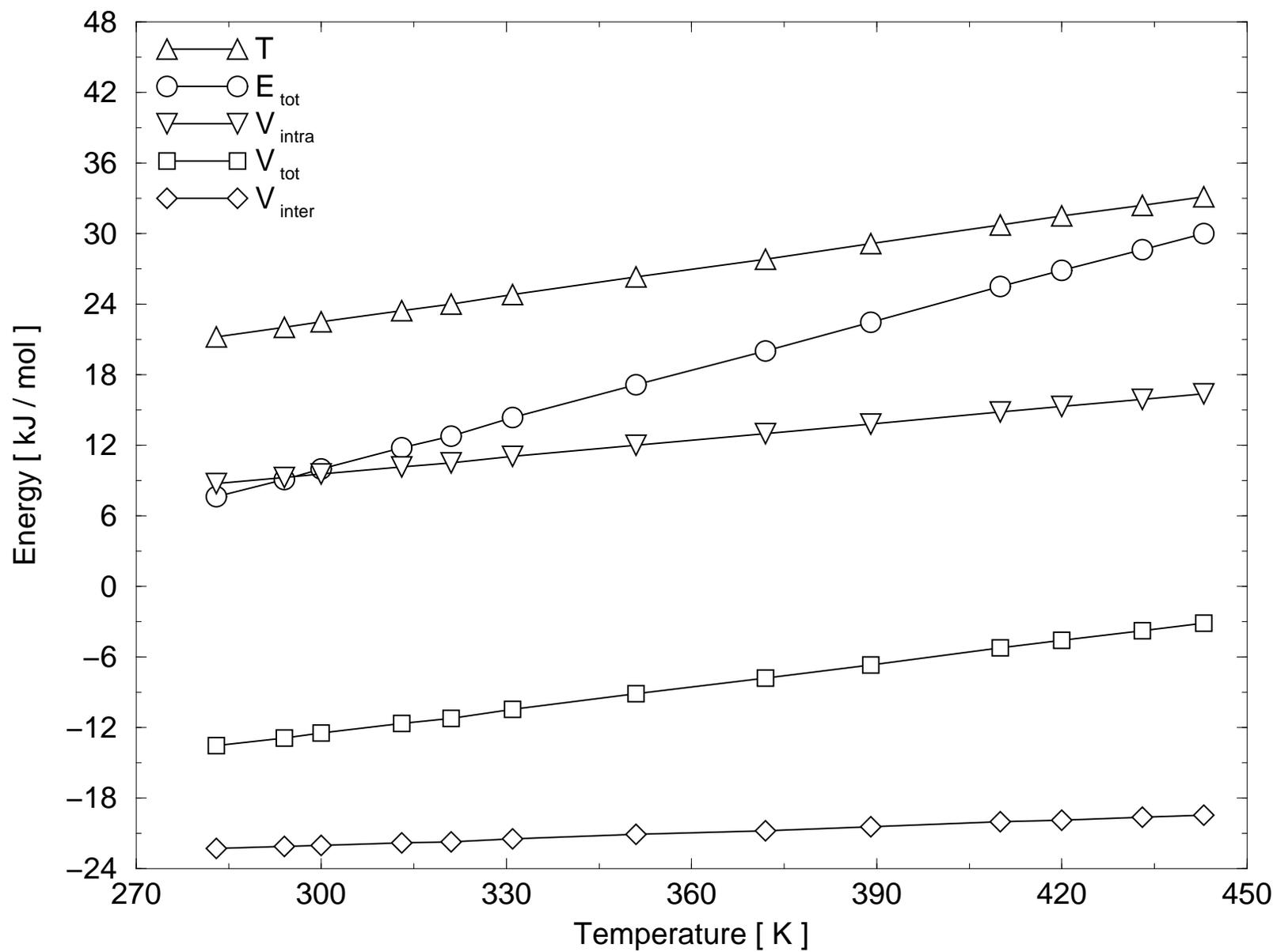
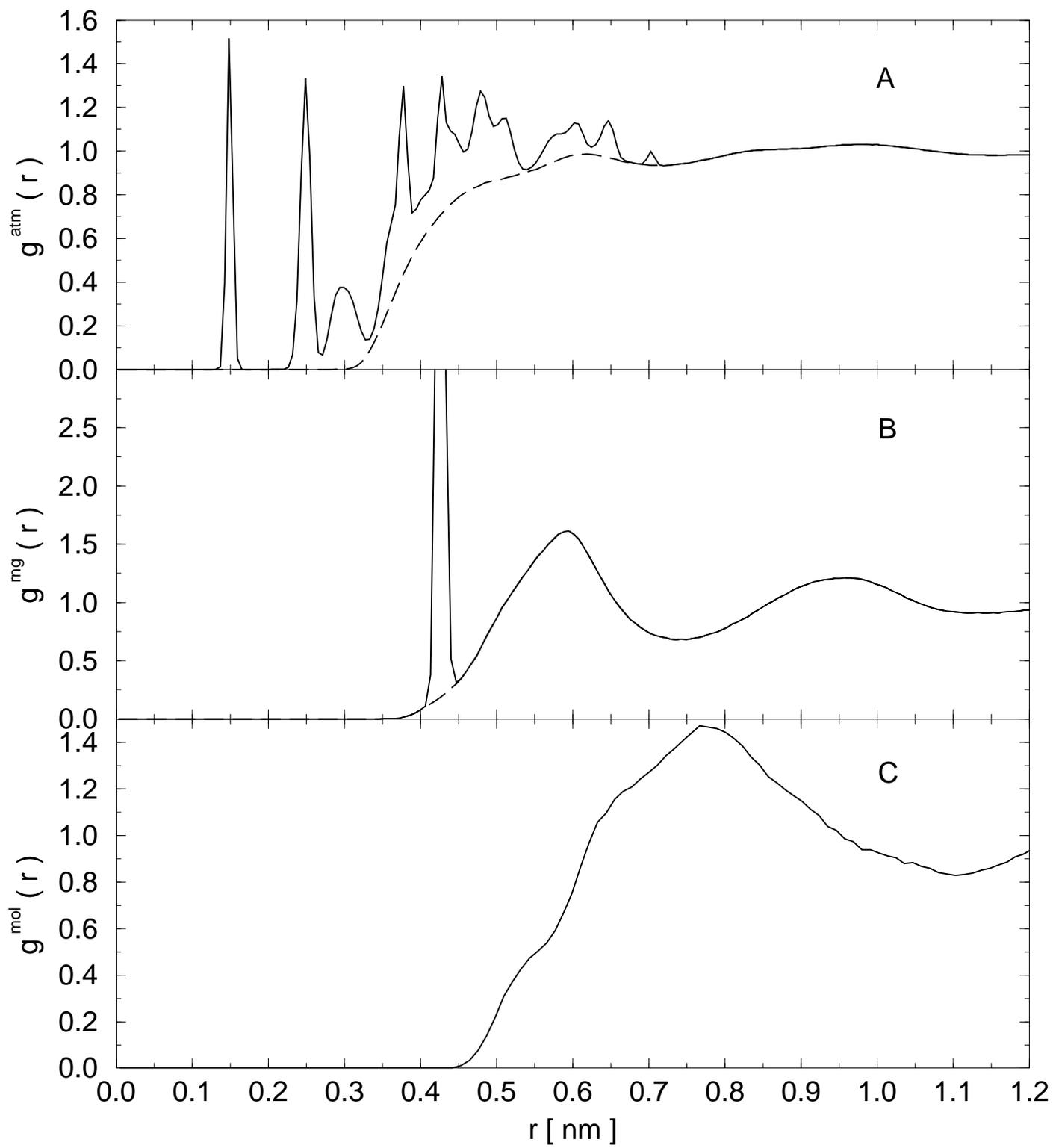
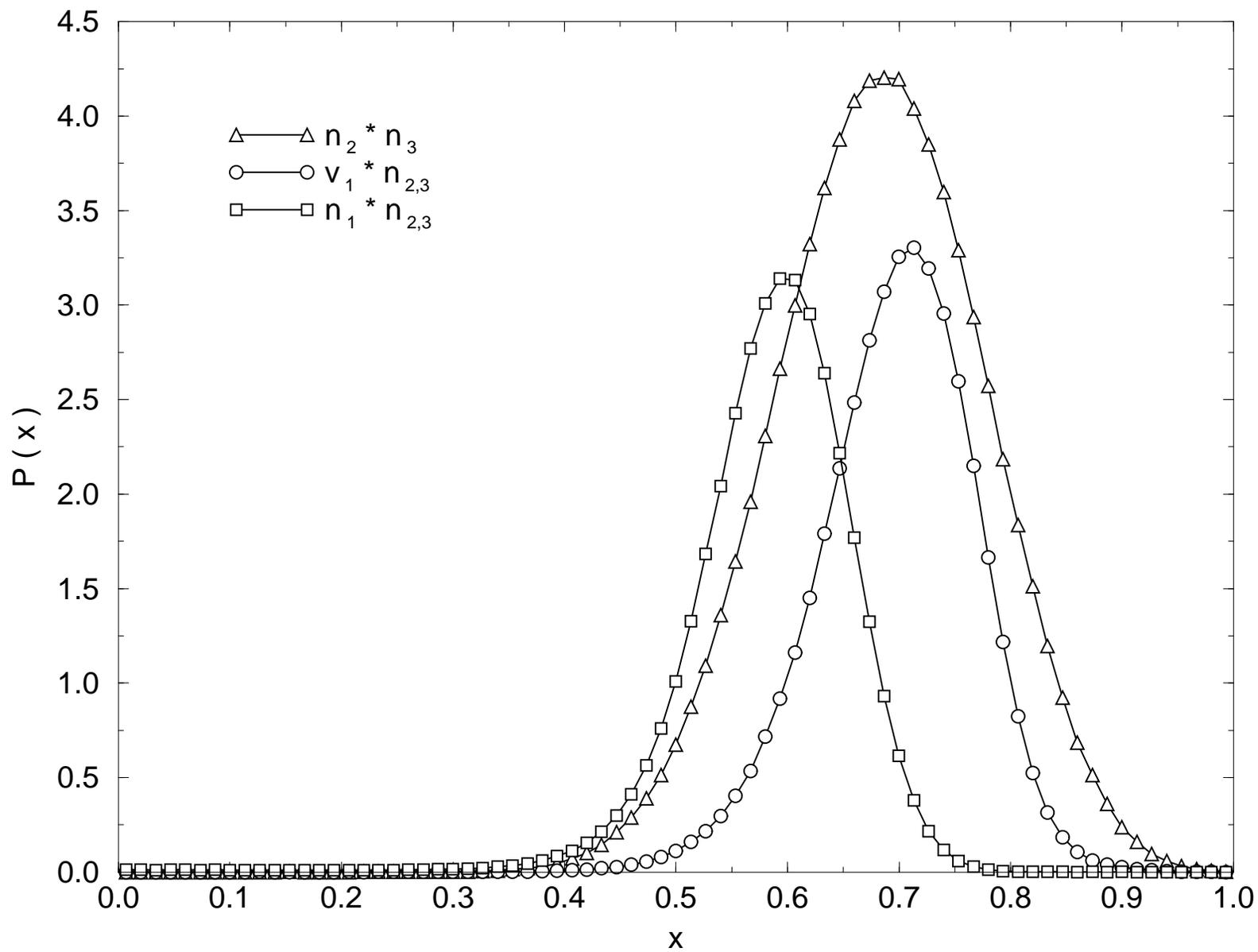
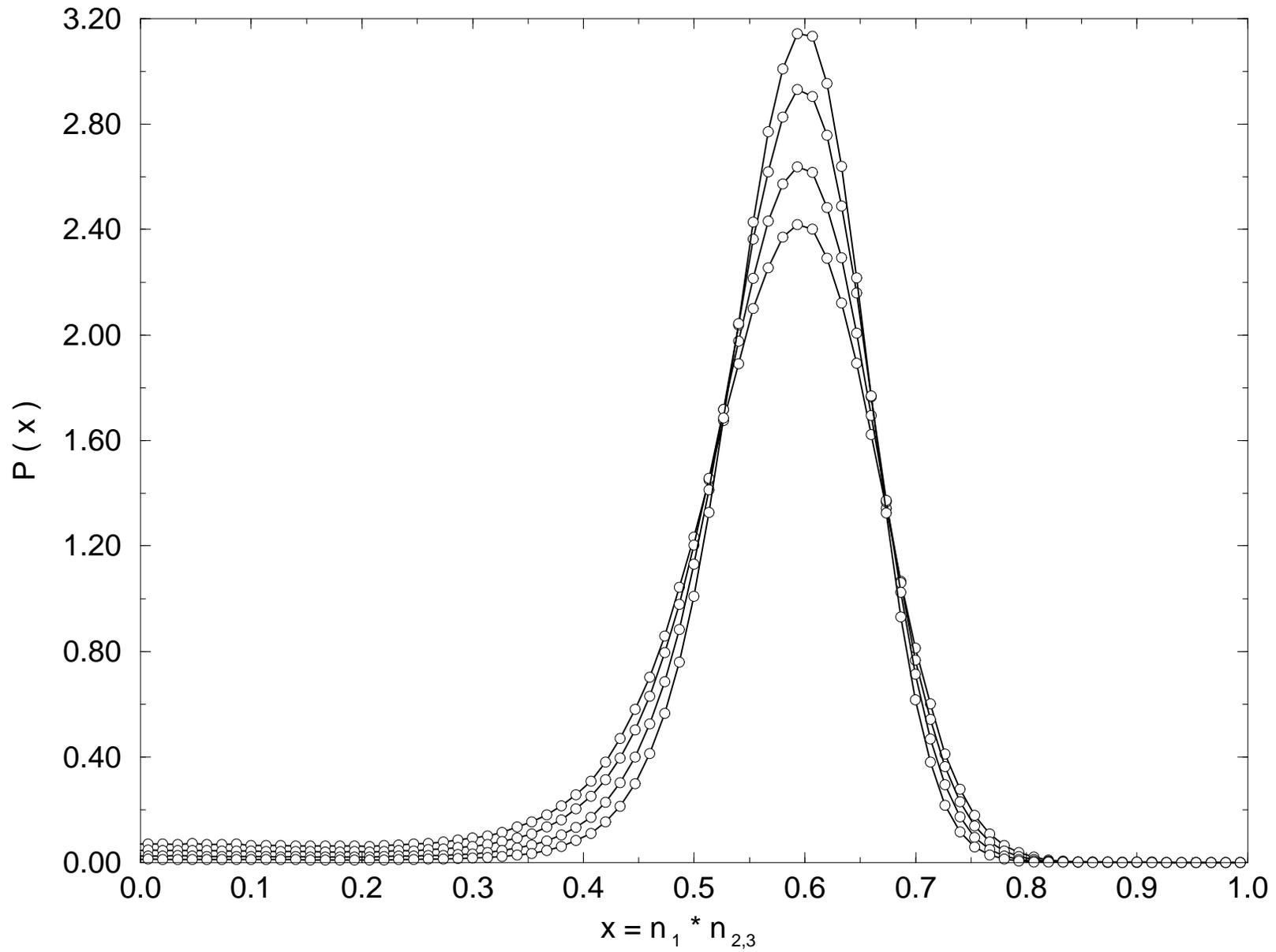
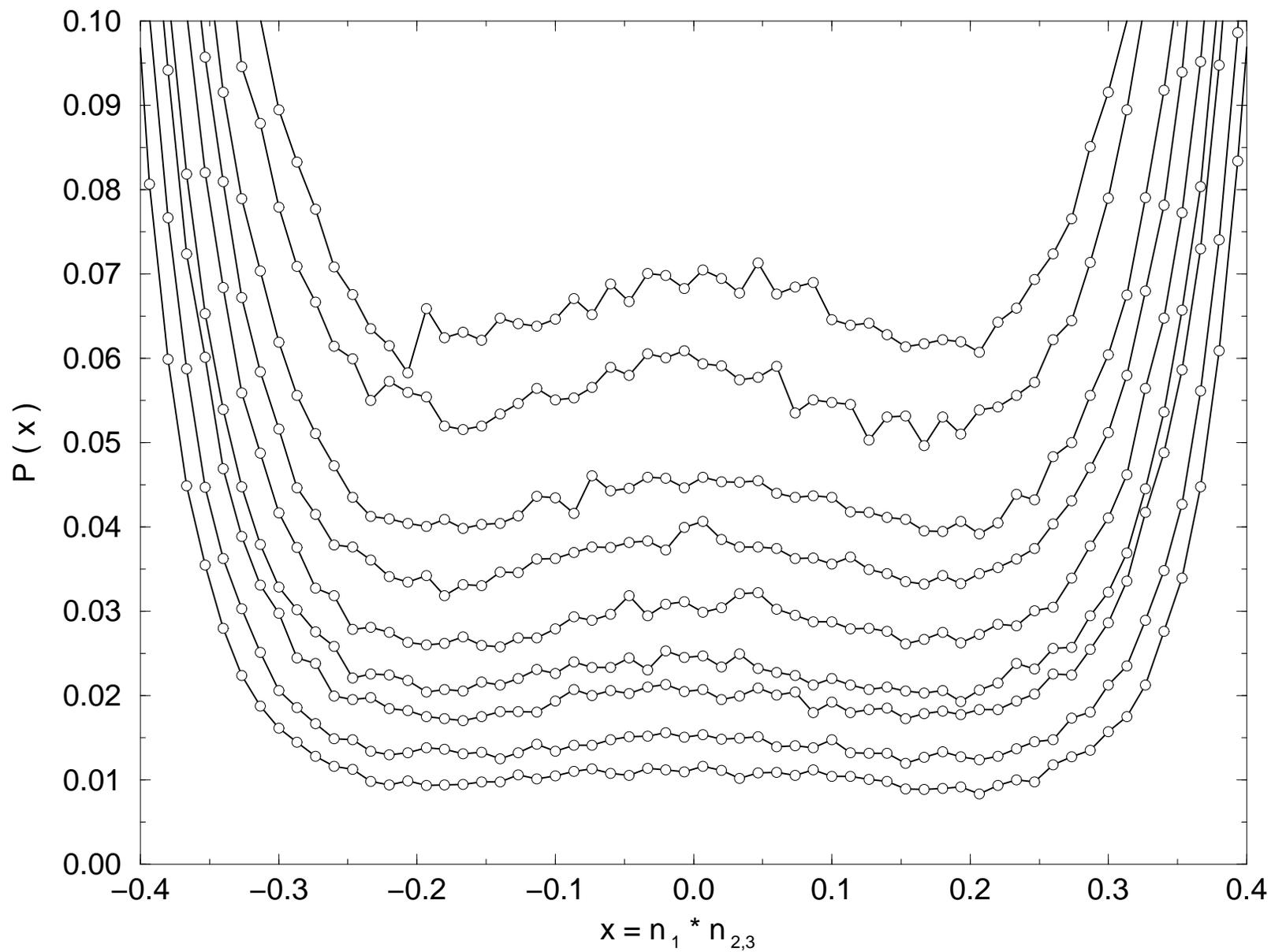
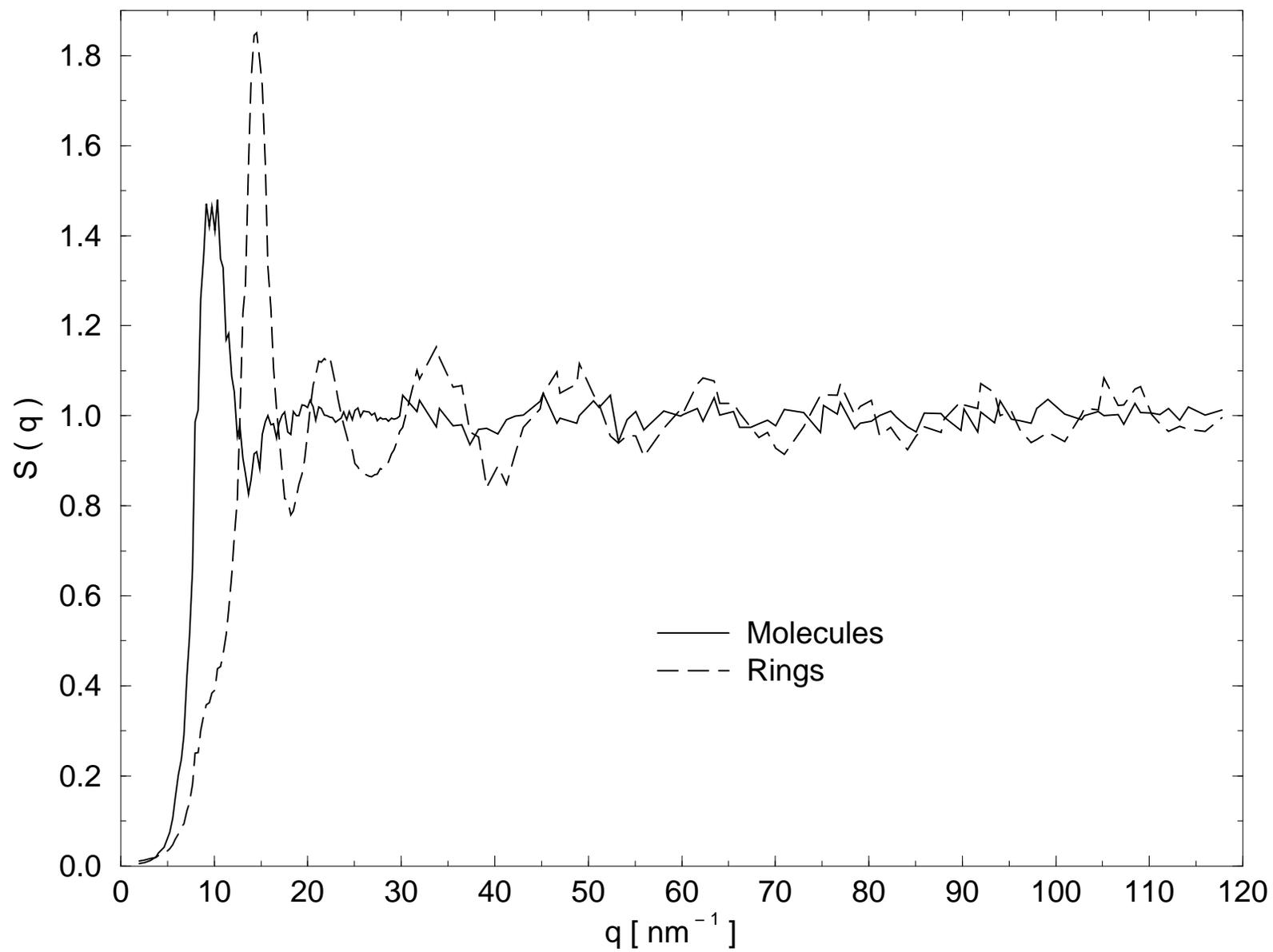
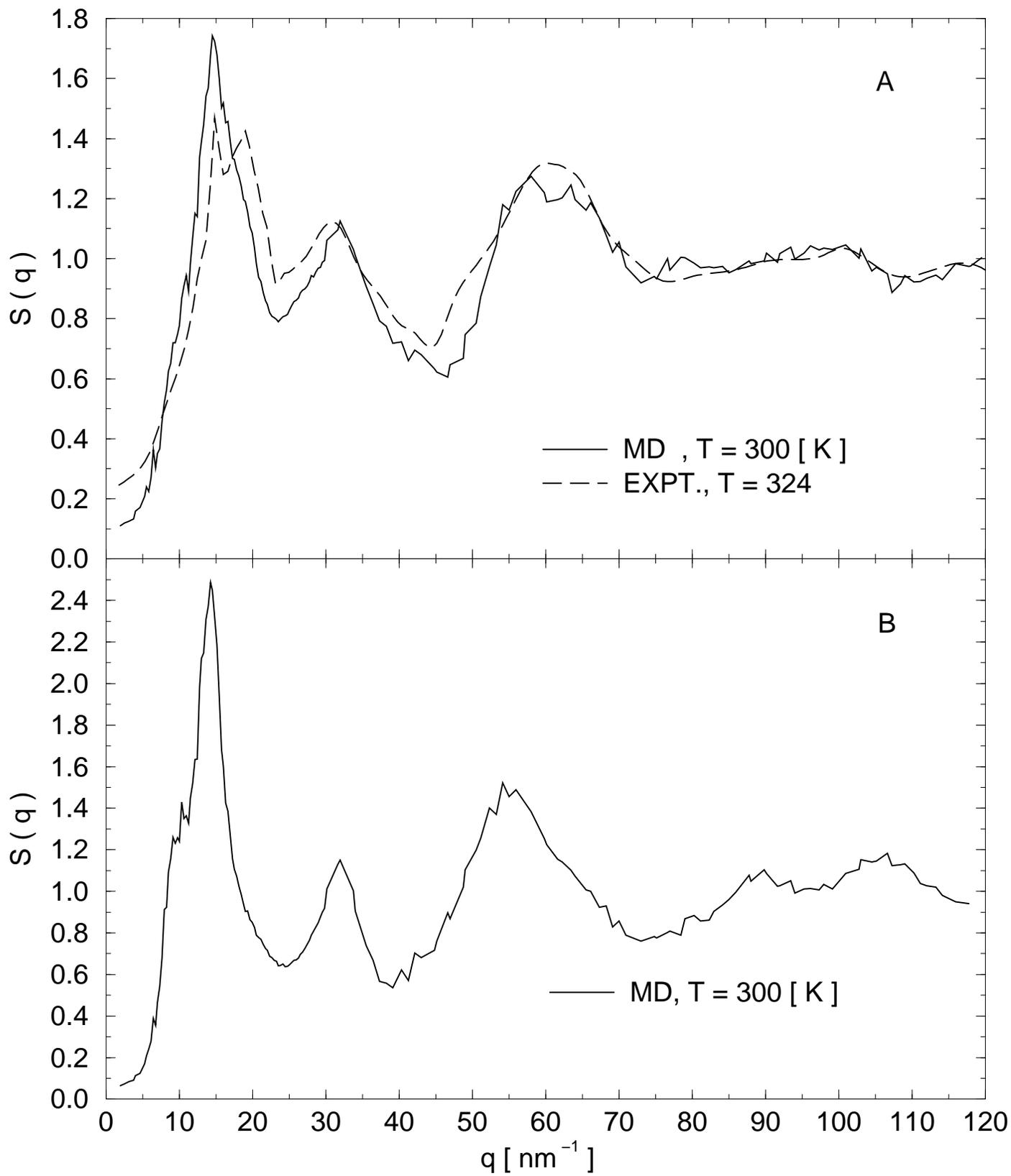
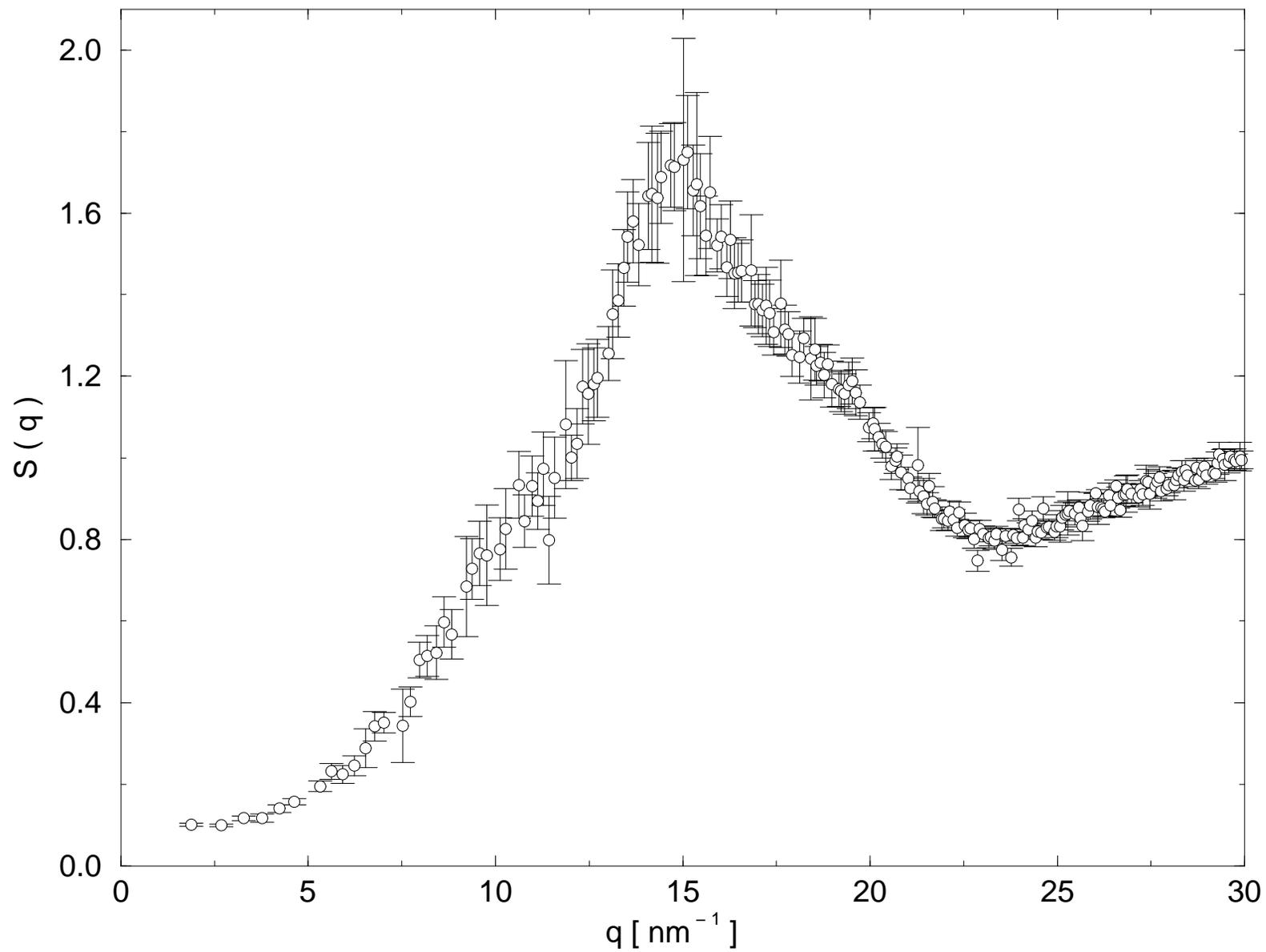
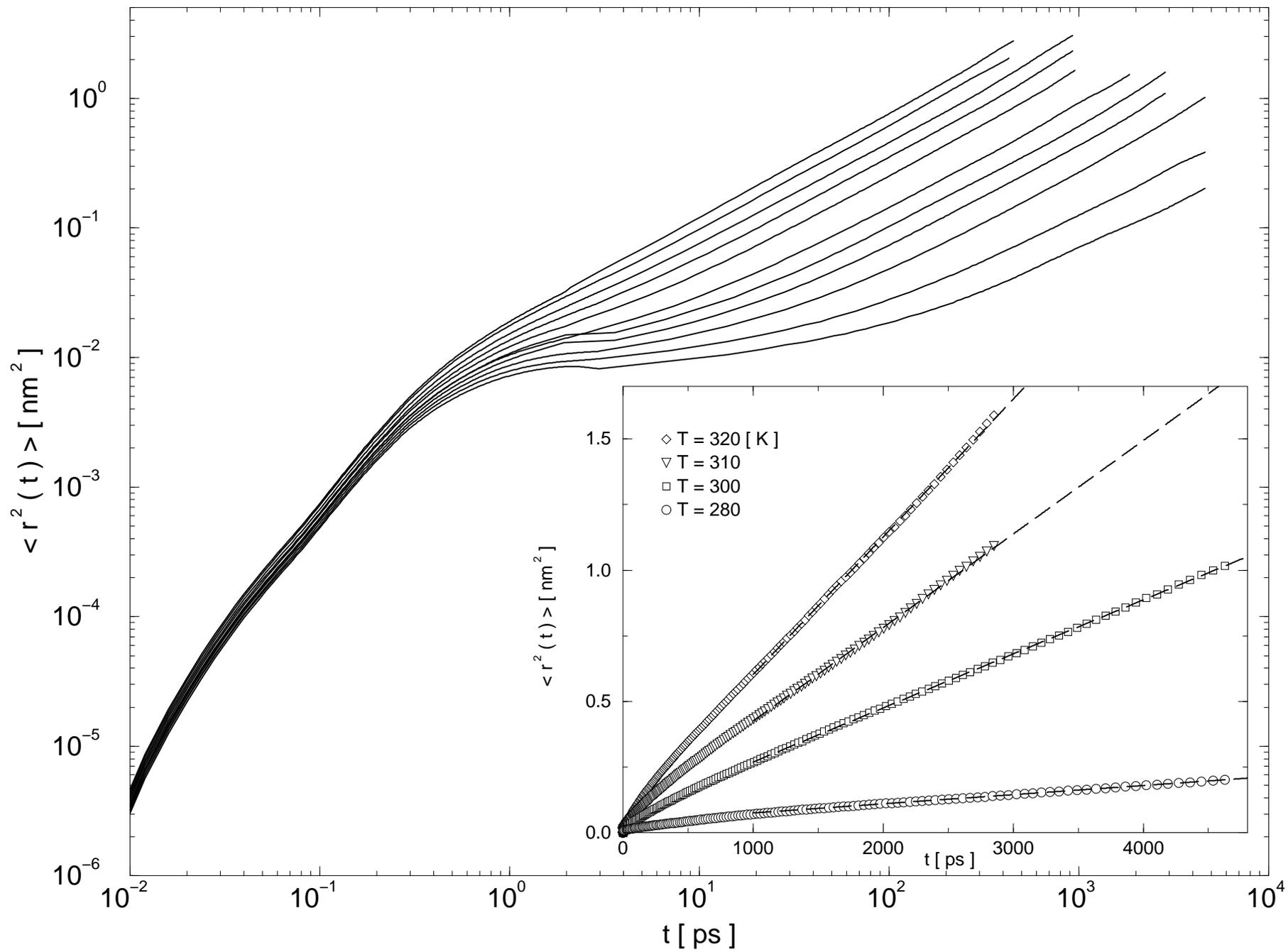
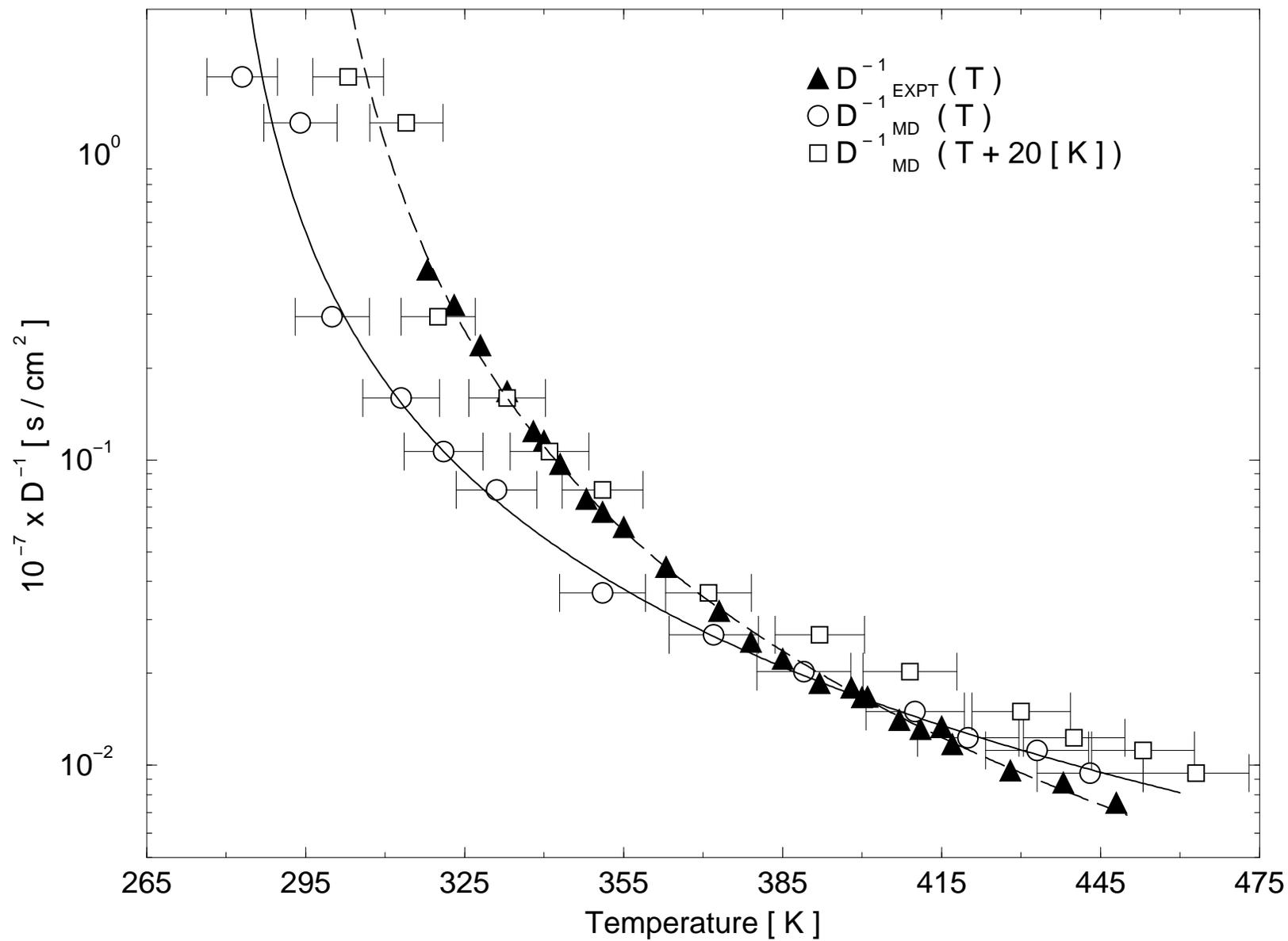
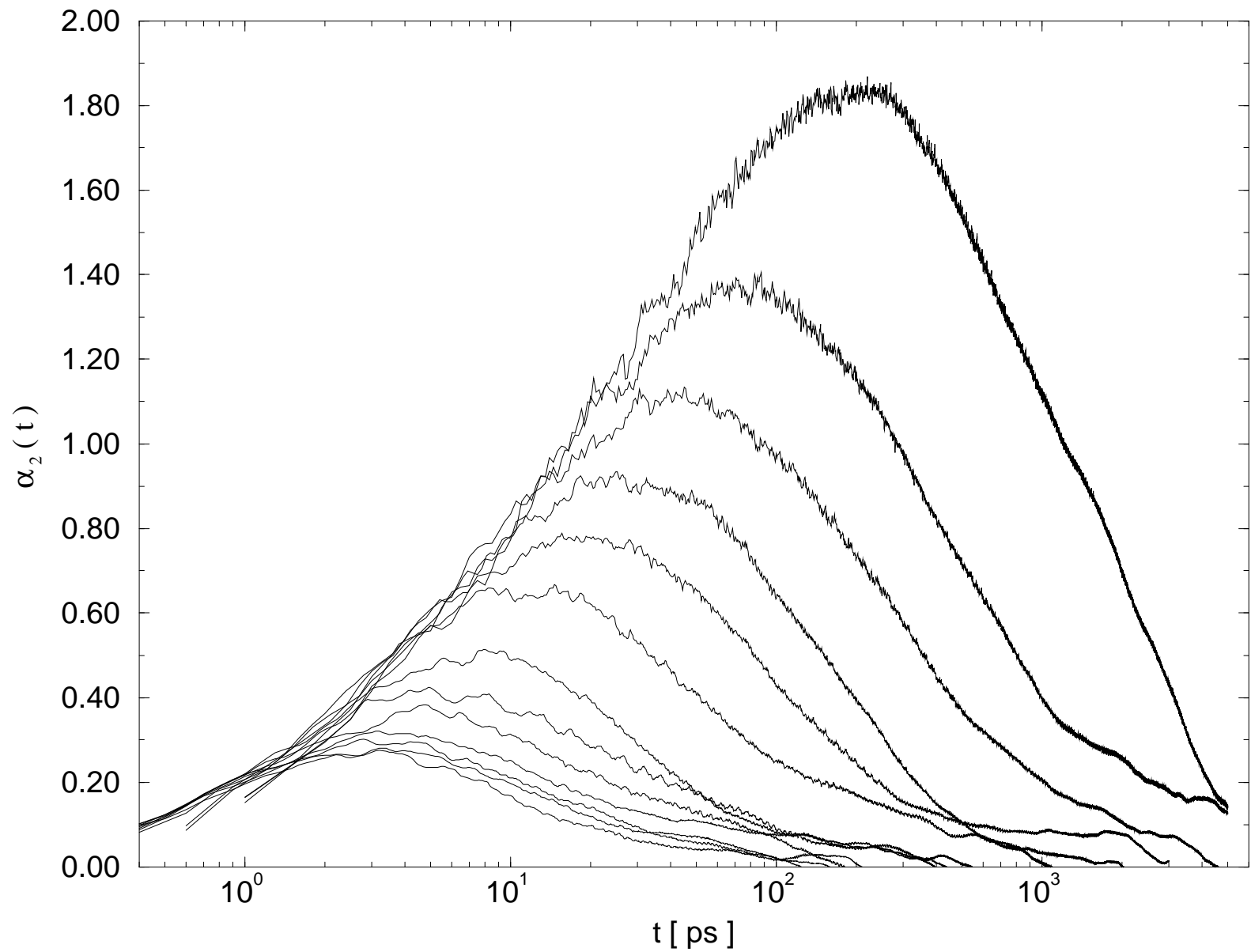
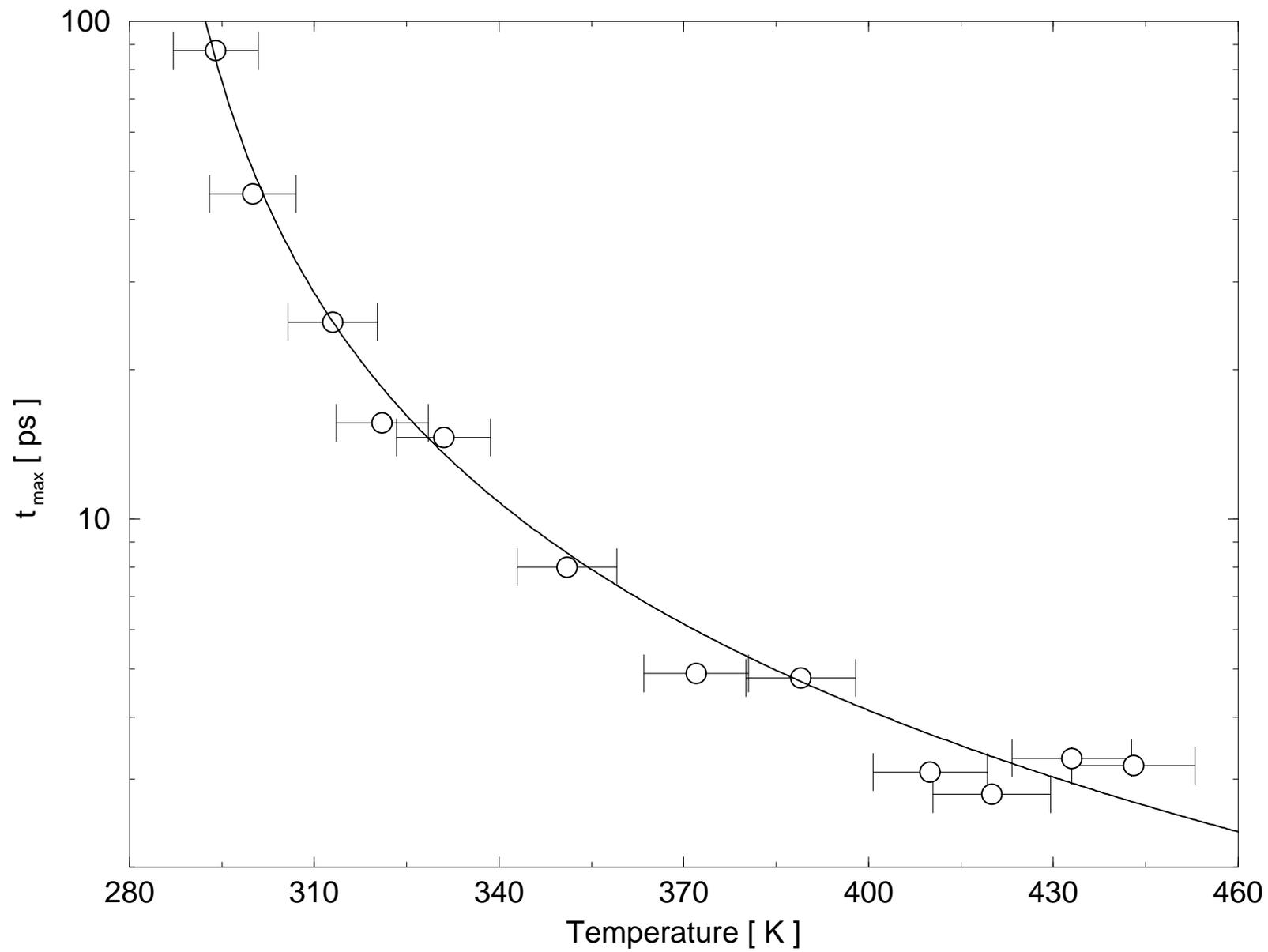
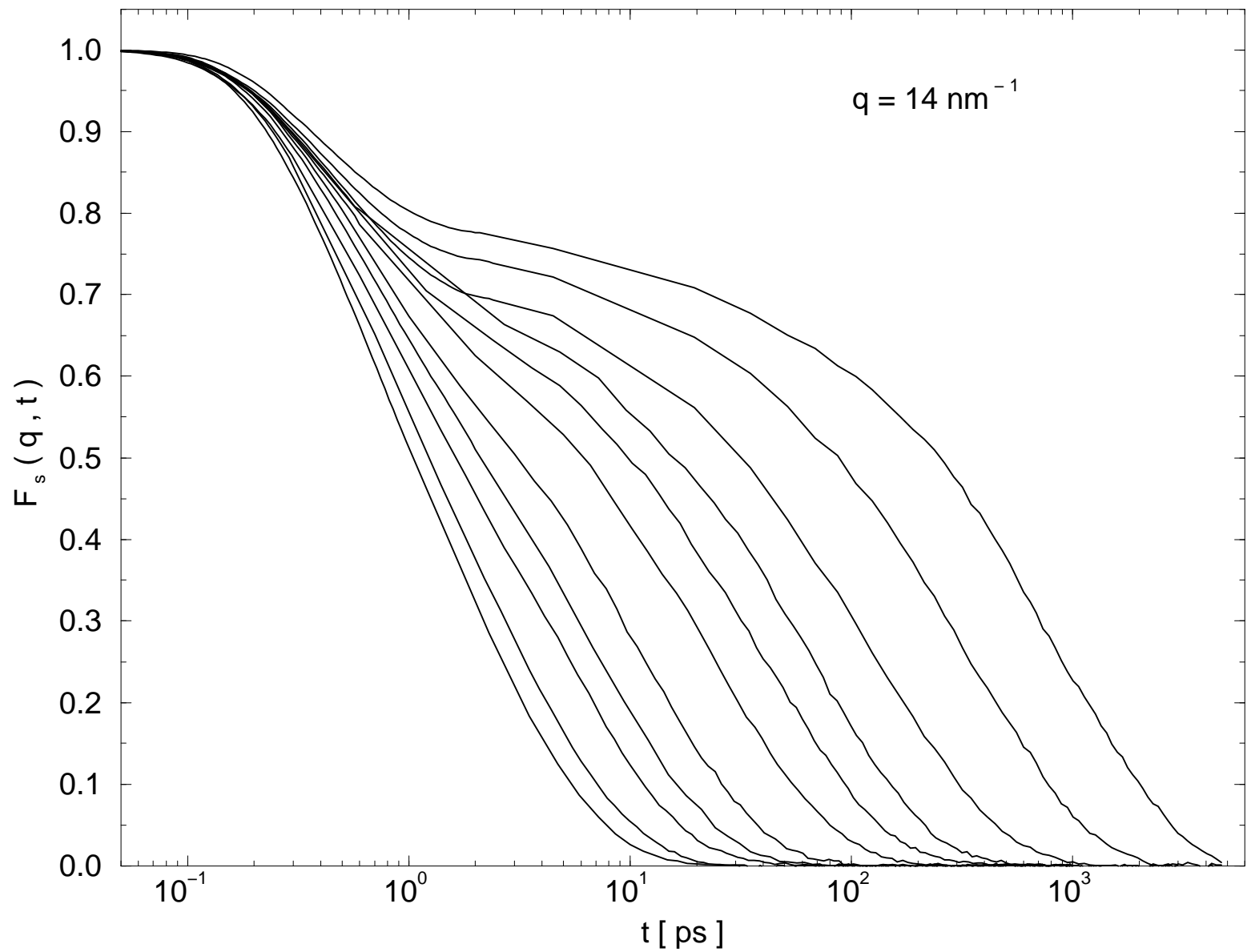
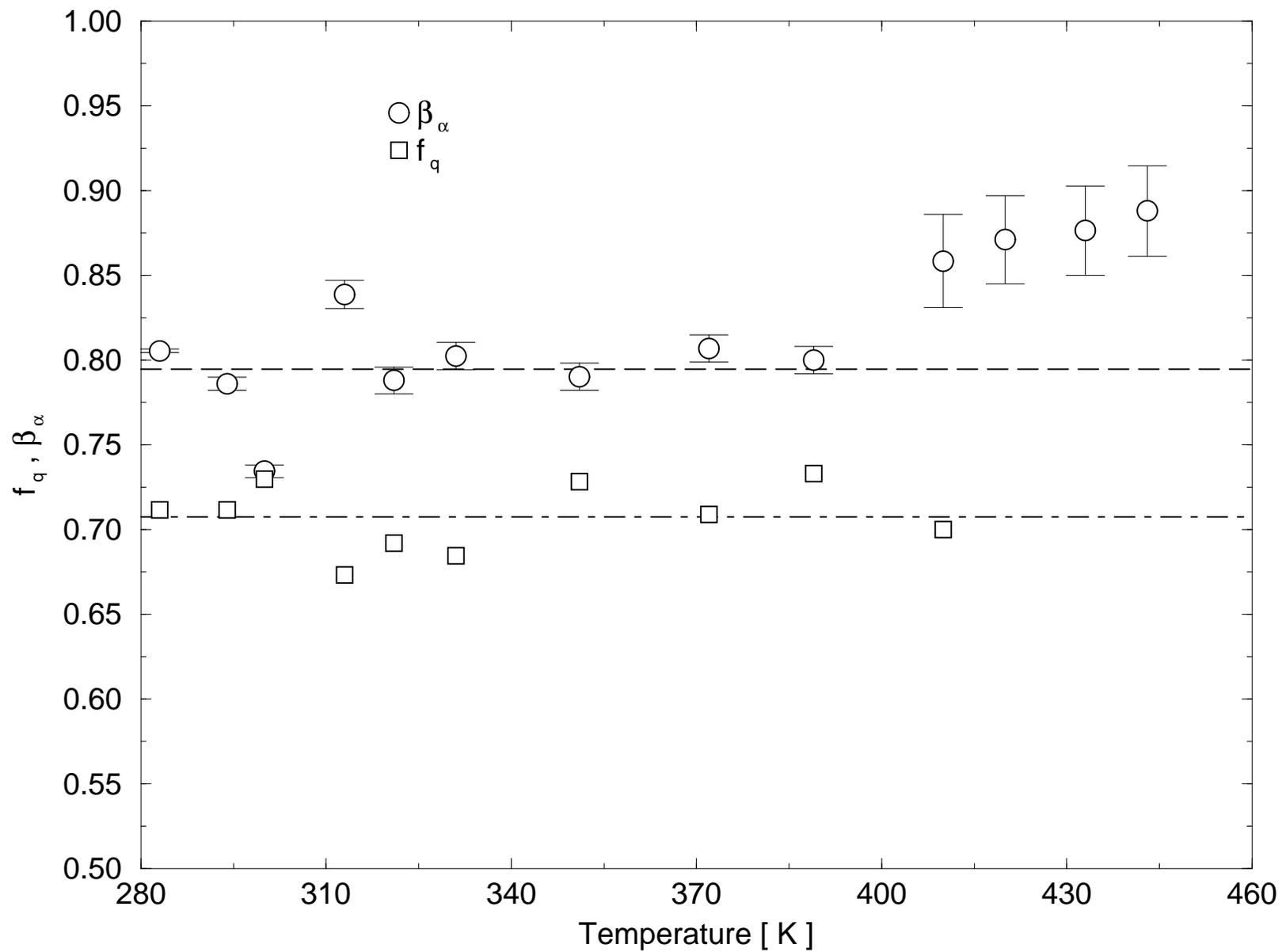
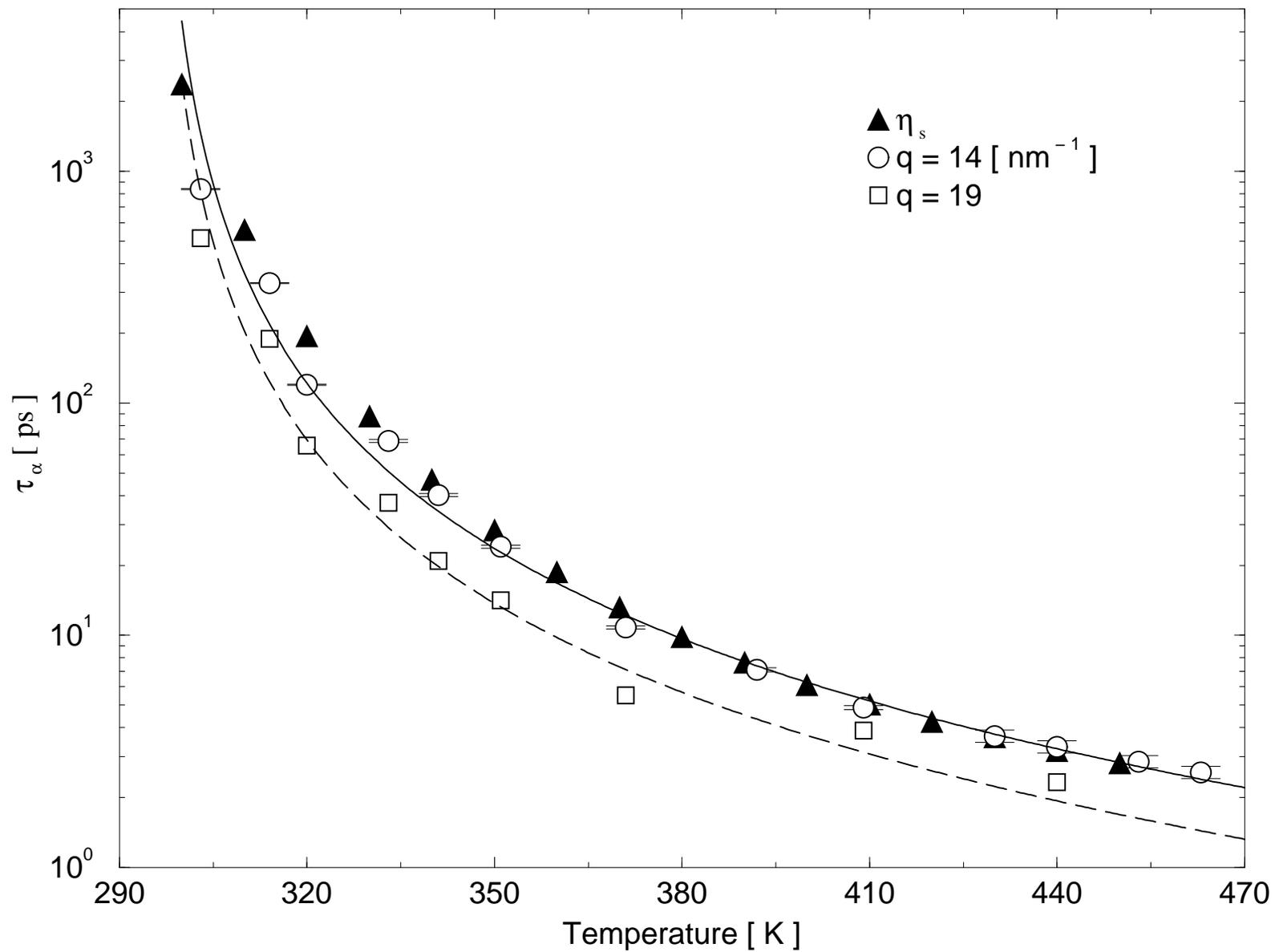
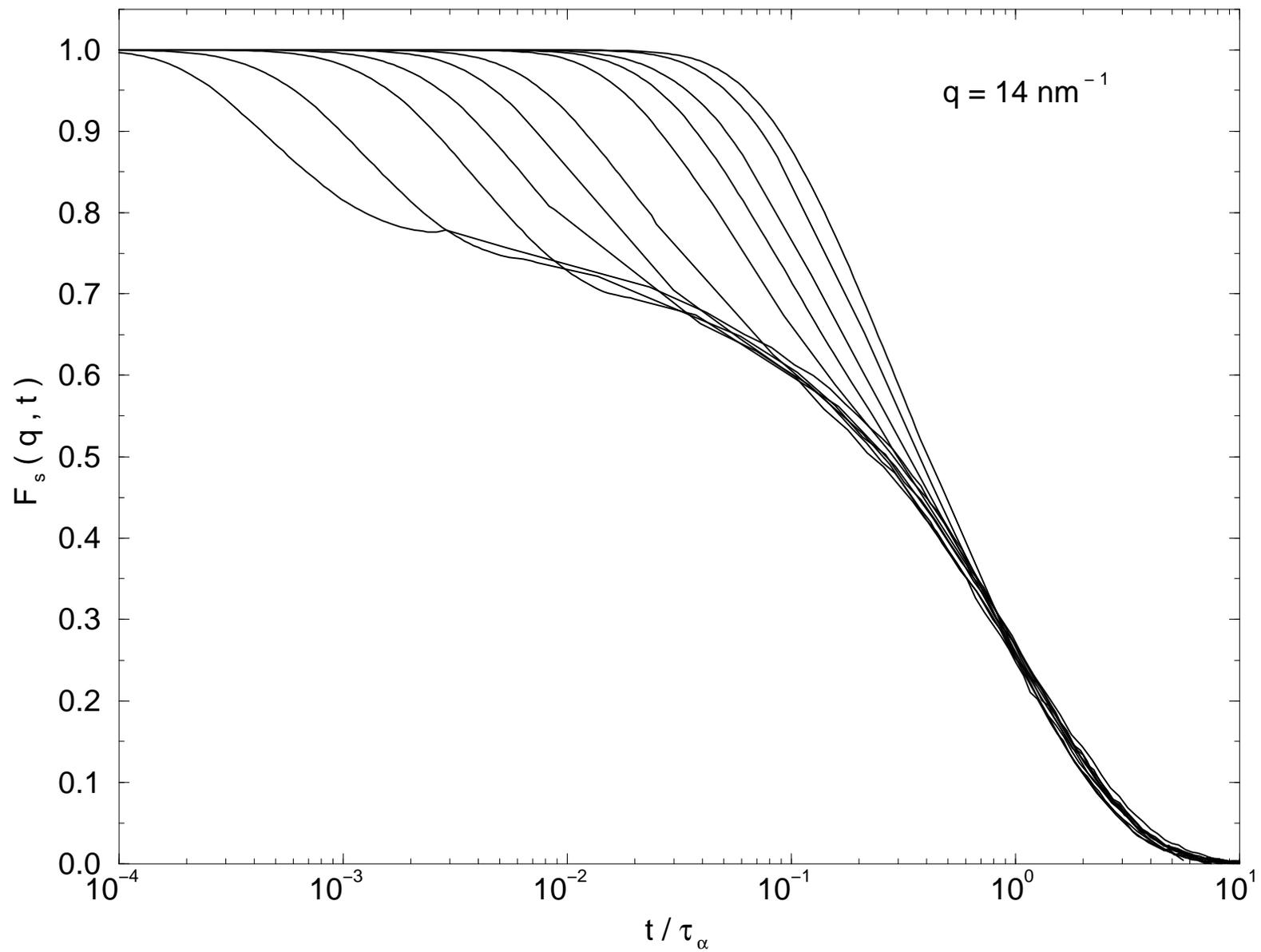
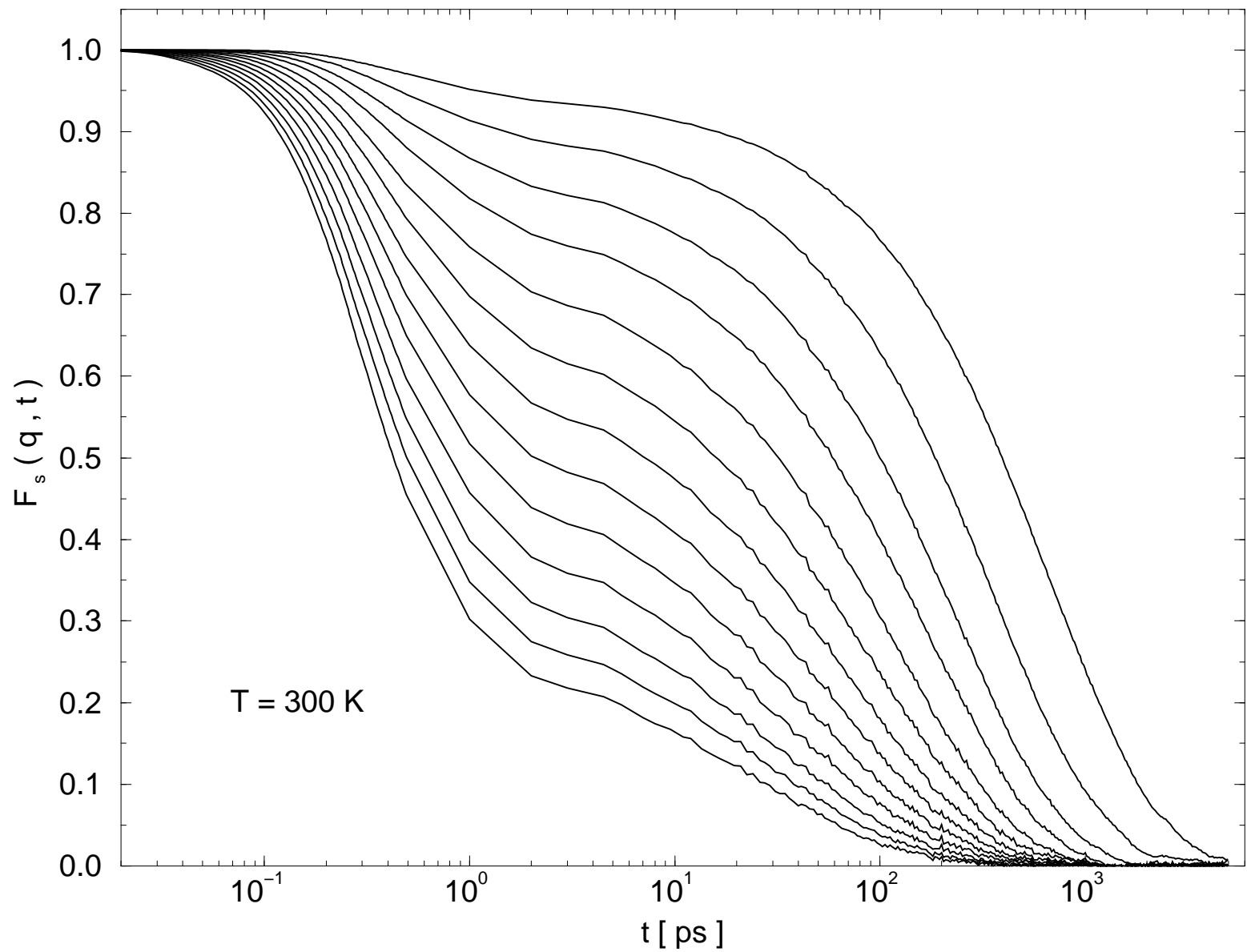
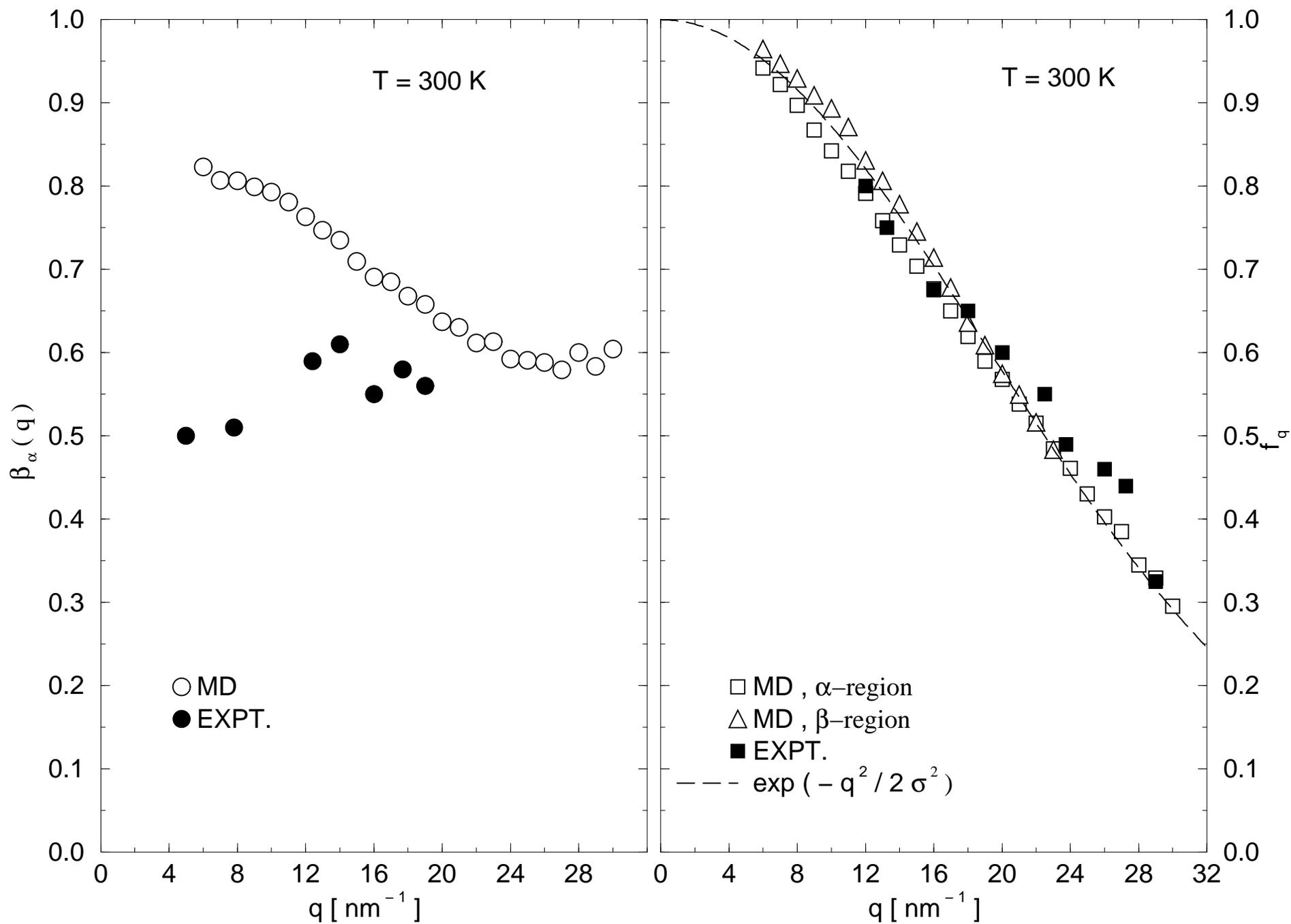
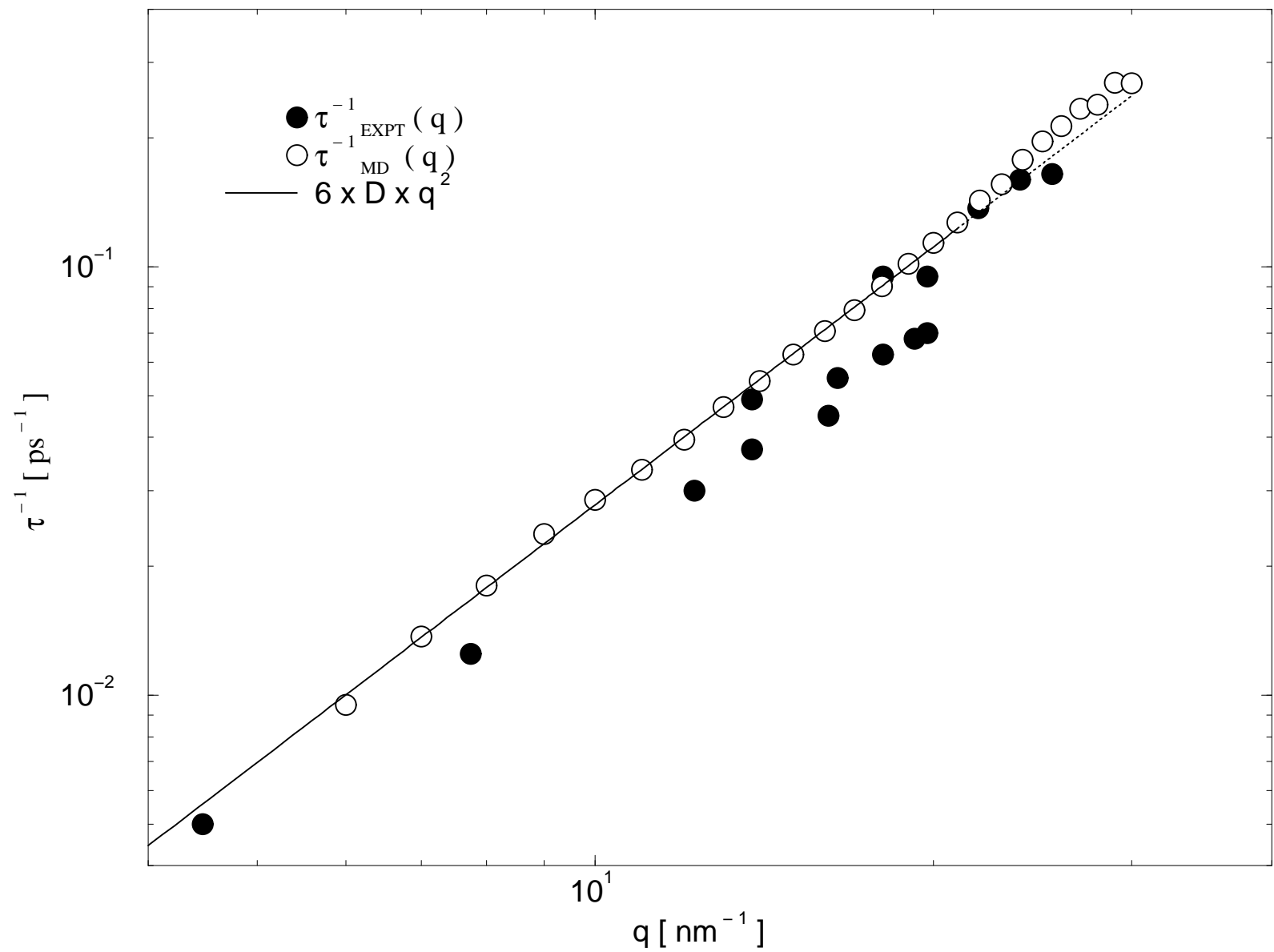
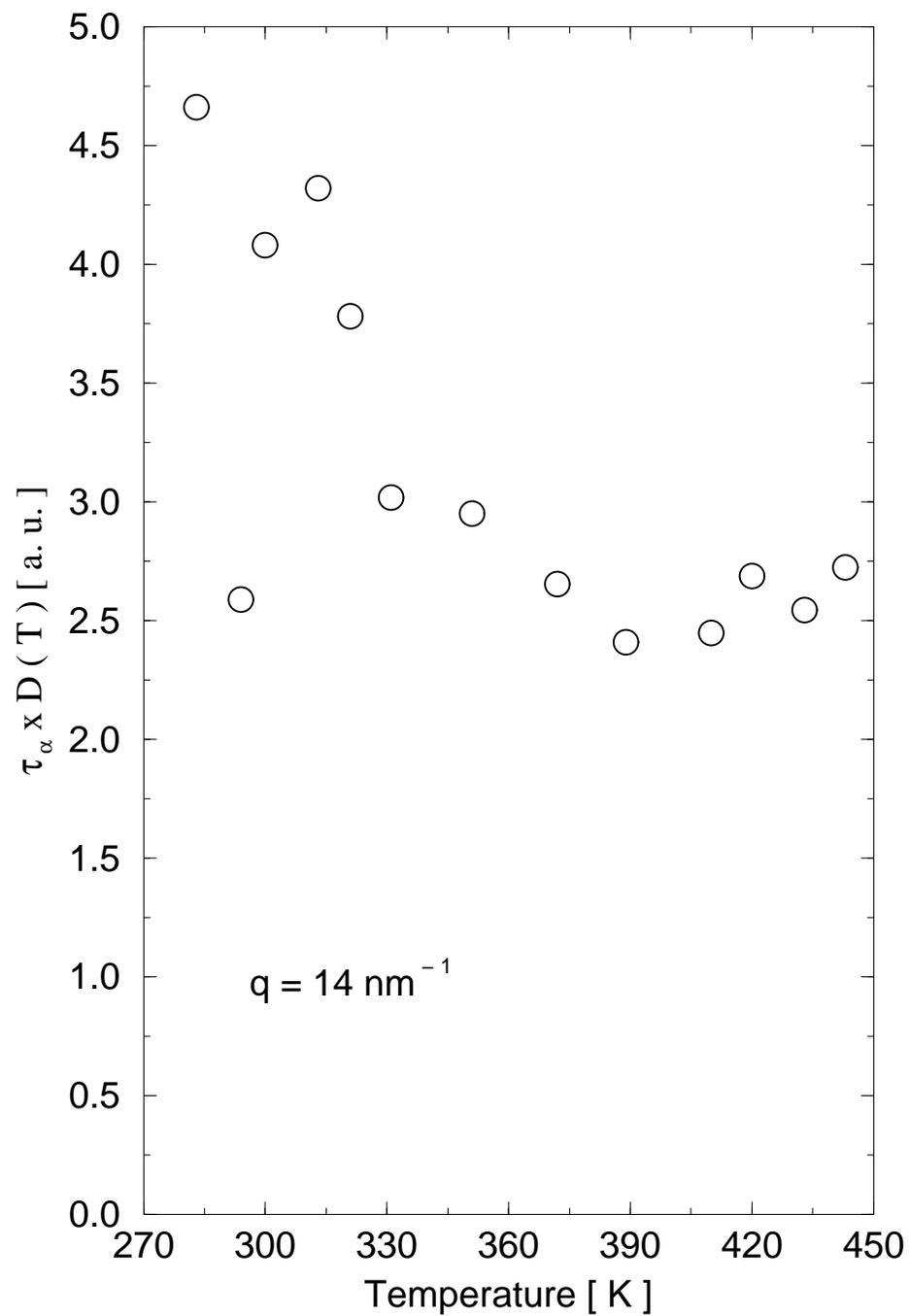
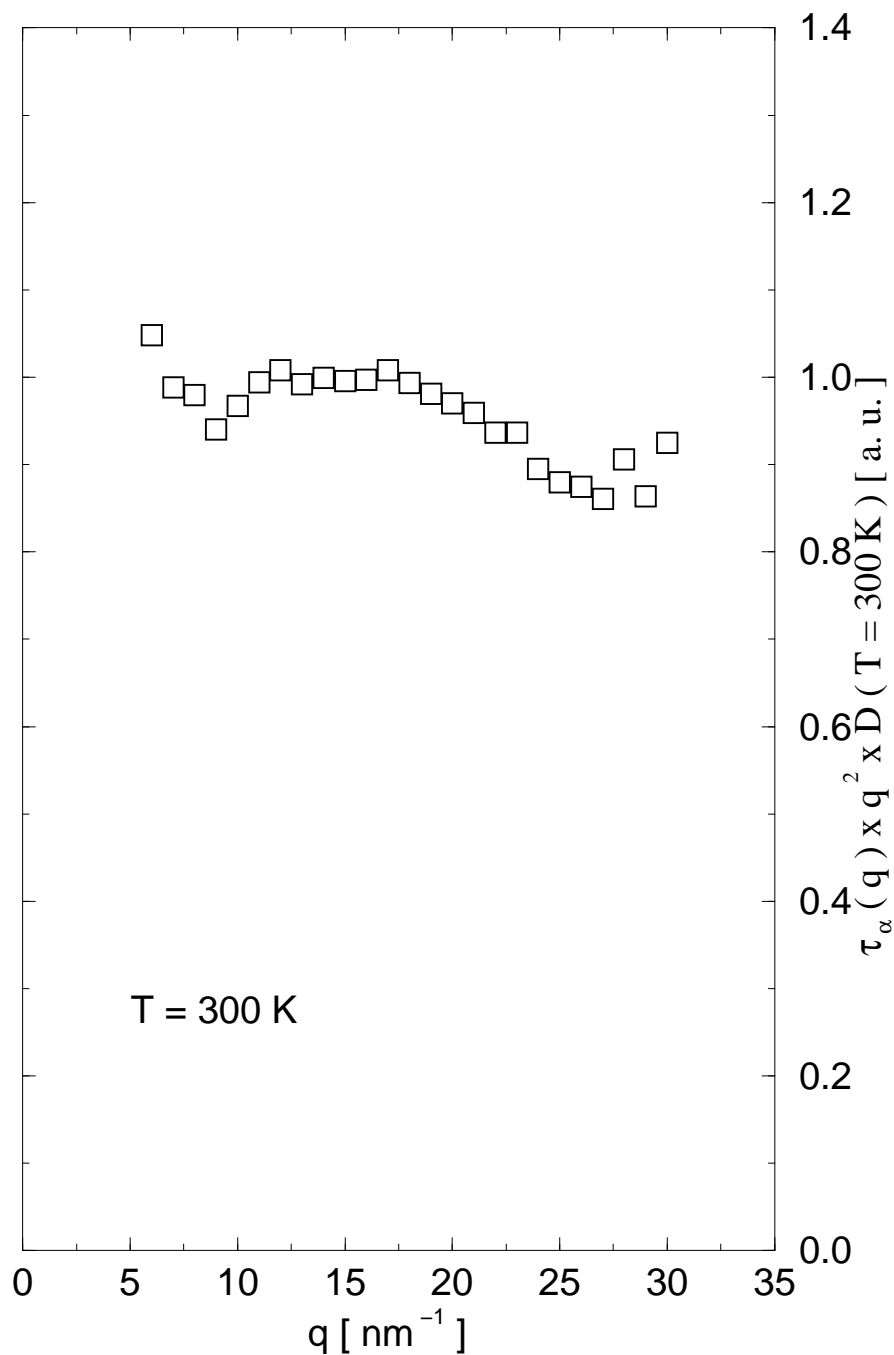
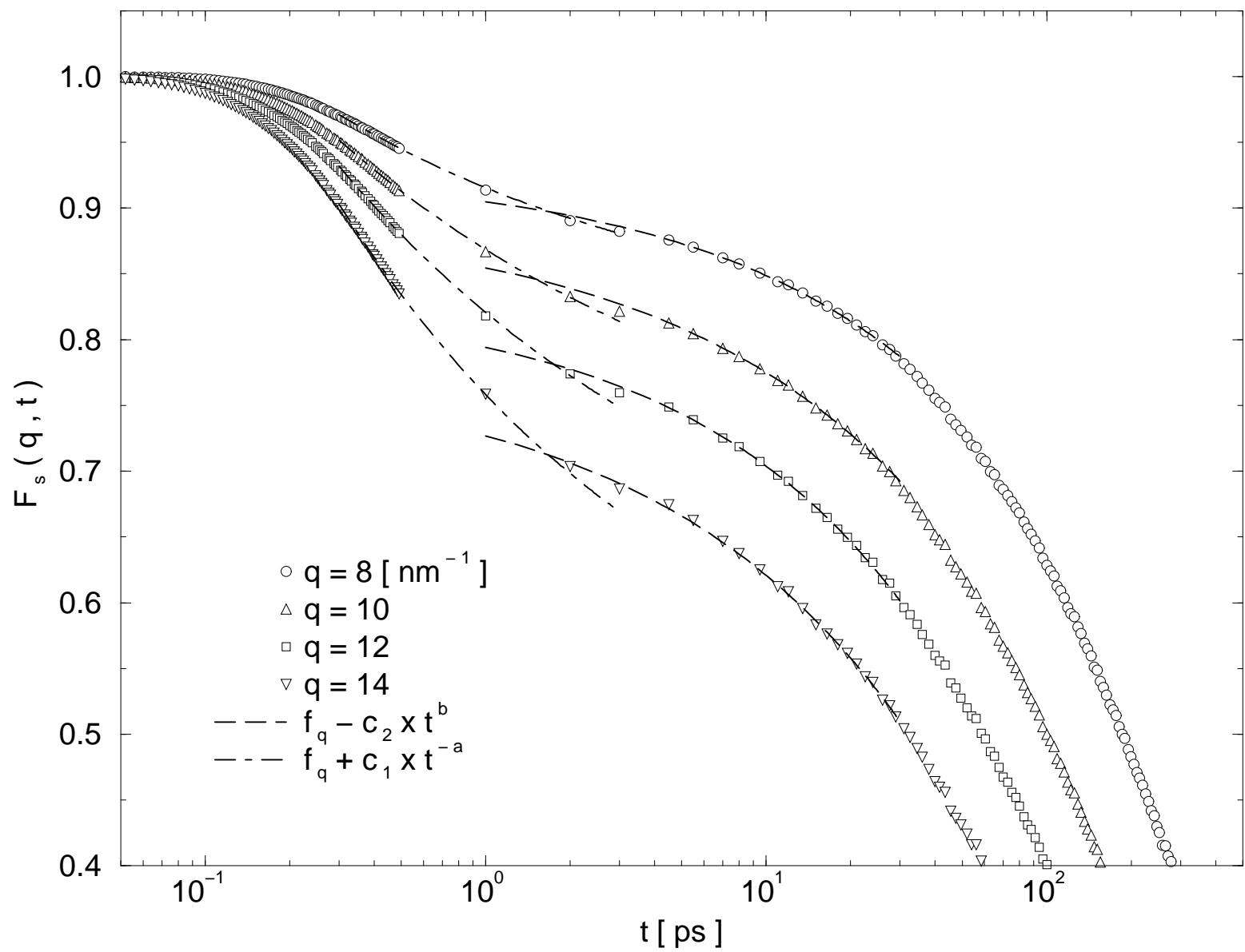
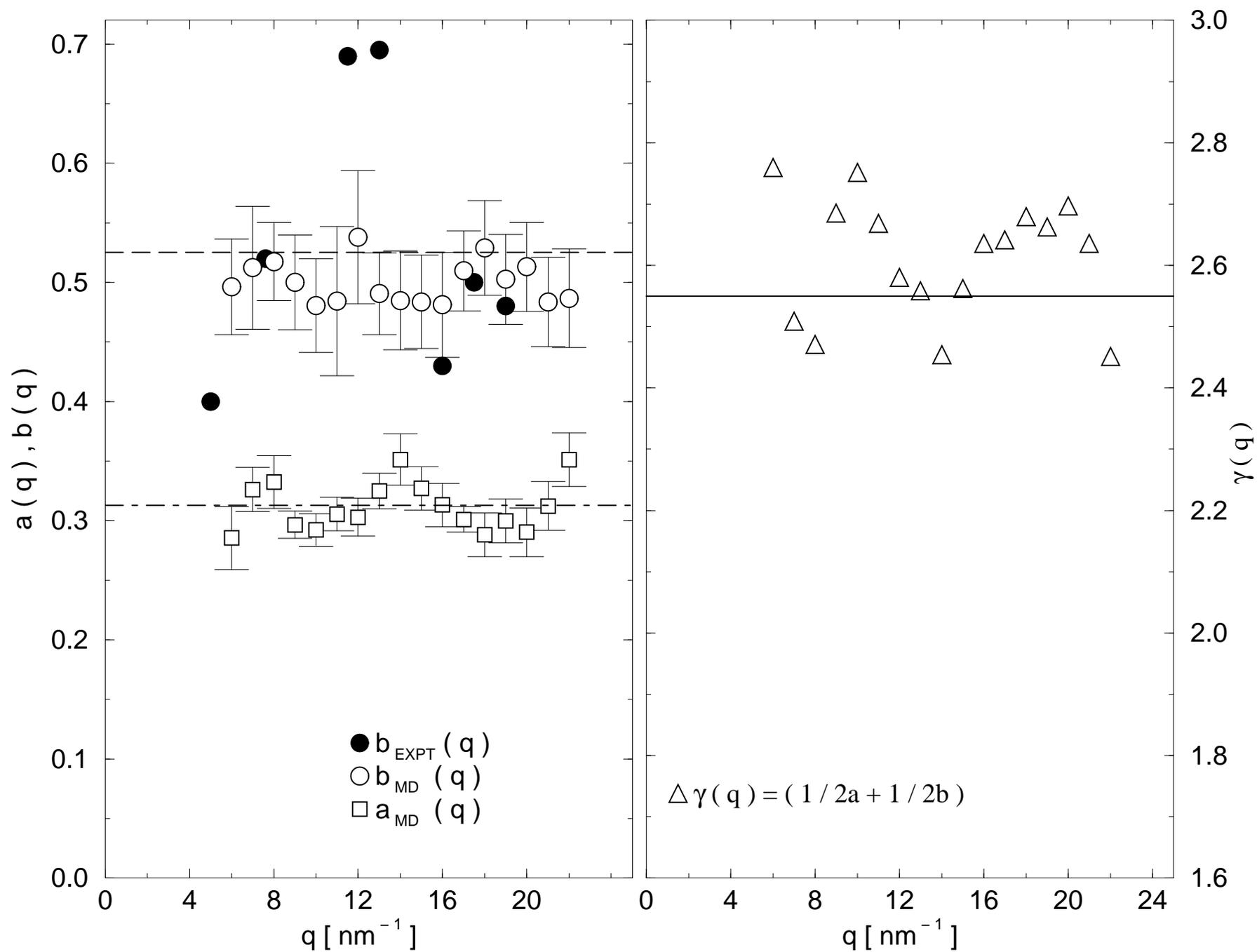
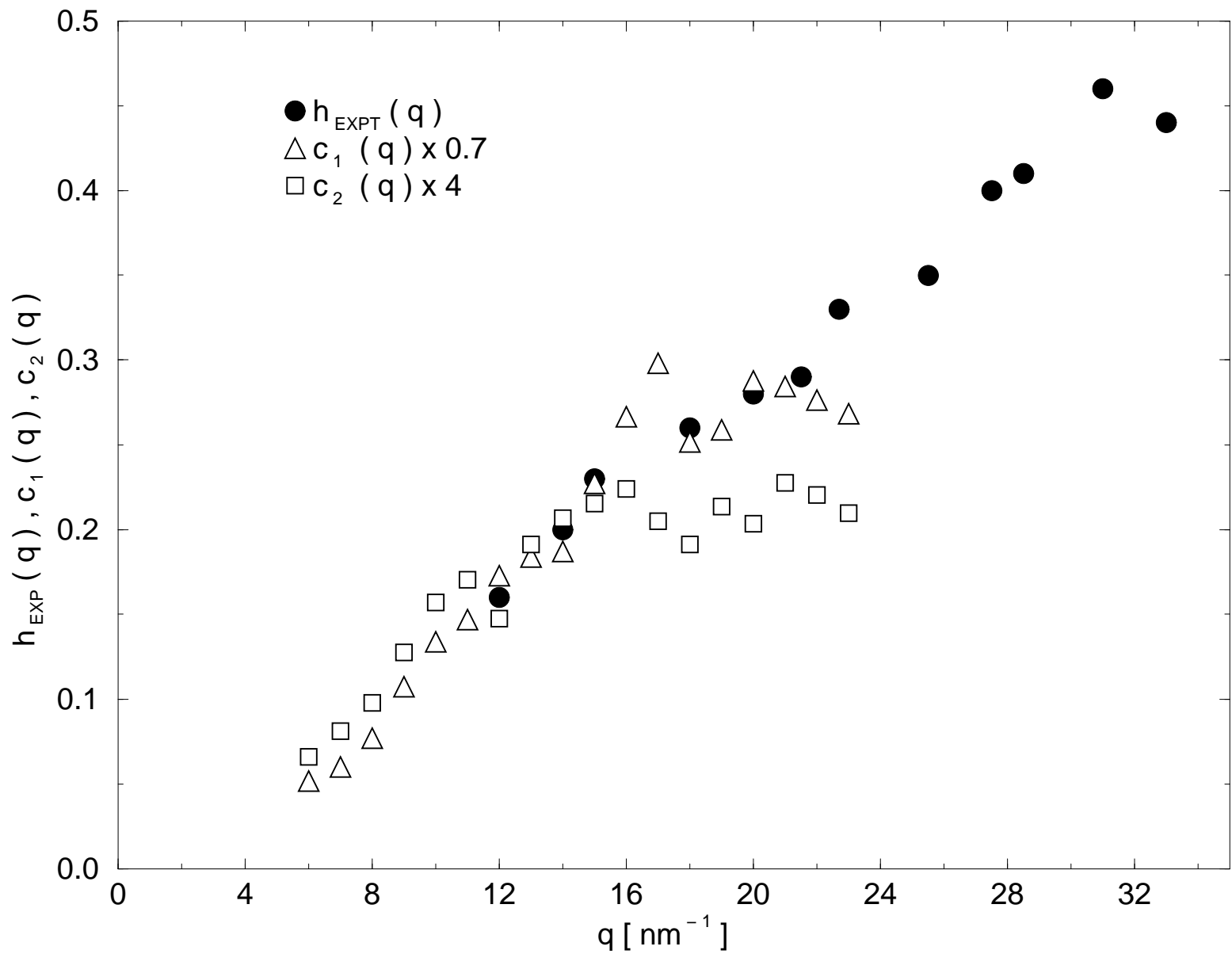
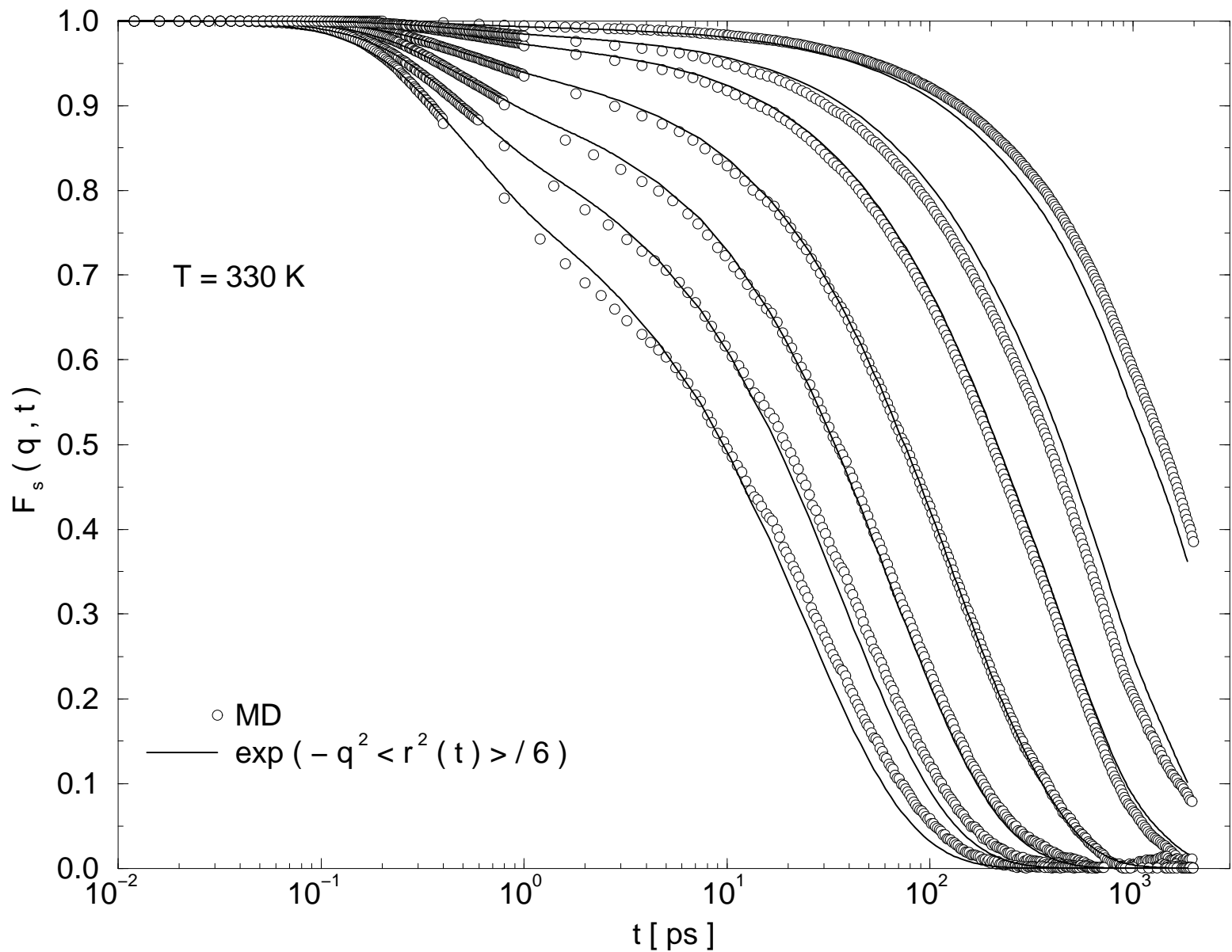
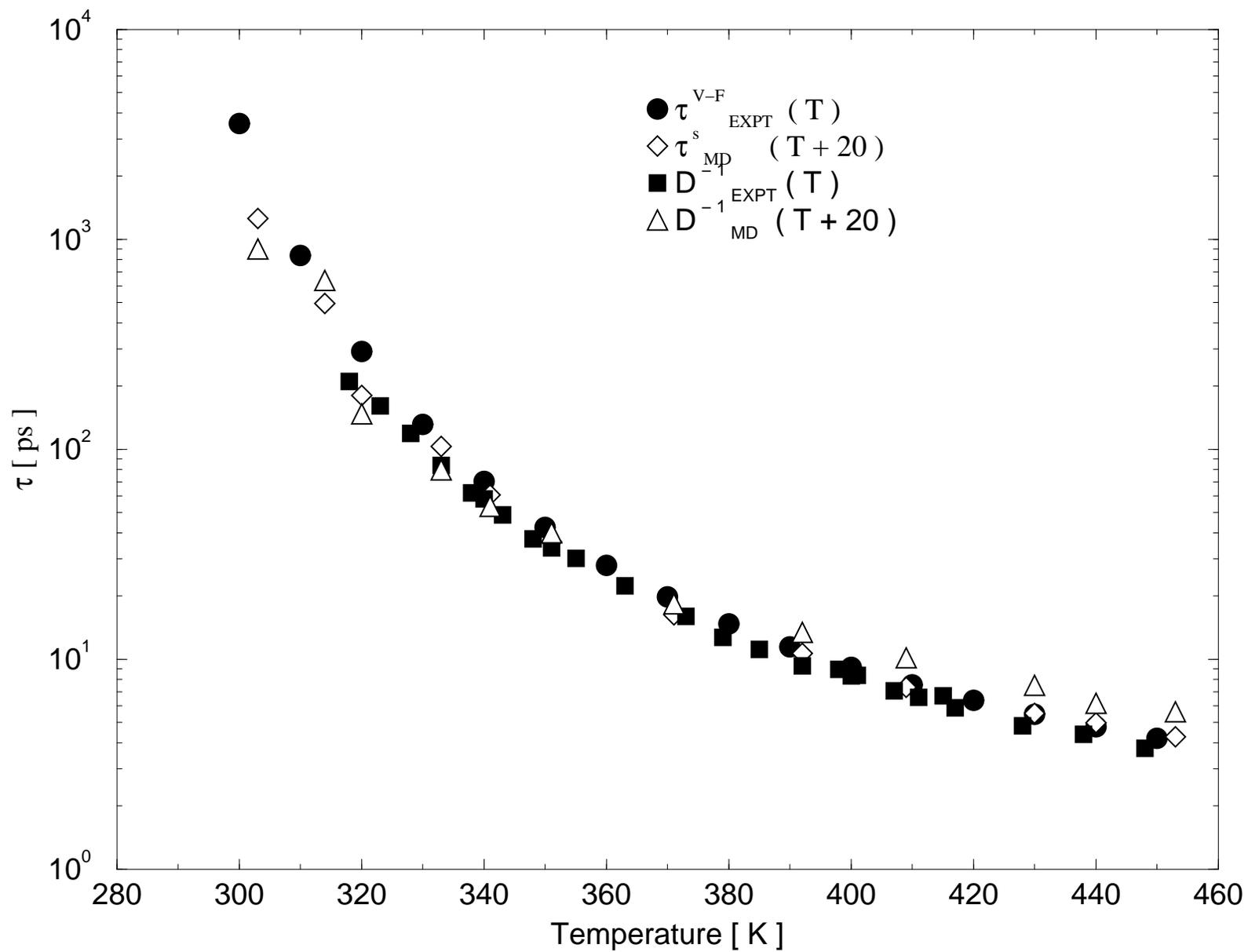
\end{document}